\def\year{2023}\relax
\documentclass[letterpaper]{article} % DO NOT CHANGE THIS
\usepackage{aaai21}  % DO NOT CHANGE THIS
\usepackage{times}  % DO NOT CHANGE THIS
\usepackage{helvet} % DO NOT CHANGE THIS
\usepackage{courier}  % DO NOT CHANGE THIS
\usepackage[hyphens]{url}  % DO NOT CHANGE THIS
\usepackage{graphicx} % DO NOT CHANGE THIS
\usepackage{natbib}  % DO NOT CHANGE THIS AND DO NOT ADD ANY OPTIONS TO IT
\usepackage{caption} % DO NOT CHANGE THIS AND DO NOT ADD ANY OPTIONS TO IT
\usepackage{siunitx}
\usepackage{subfig}
\usepackage{placeins}
\usepackage{stfloats}

\usepackage[dvipsnames,usenames]{xcolor}

\def\capfirstletteraux#1#2\relax{\uppercase{#1}\lowercase{#2}}

\newcommand{\emo}[1]{\textsf{\textsc{#1}}}

\newcommand{\emoanger}[1]{{\emo{\vphantom{g}#1}}}
\newcommand{\emofear}[1]{{\emo{\vphantom{g}#1}}}
\newcommand{\emoanticipation}[1]{{\emo{\vphantom{g}#1}}}
\newcommand{\emosurprise}[1]{{\emo{\vphantom{g}#1}}}
\newcommand{\emojoy}[1]{{\emo{\vphantom{g}#1}}}
\newcommand{\emosadness}[1]{{\emo{\vphantom{g}#1}}}
\newcommand{\emotrust}[1]{{\emo{\vphantom{g}#1}}}
\newcommand{\emodisgust}[1]{{\emo{\vphantom{g}#1}}}

\newcommand{\emodisapproval}[1]{{\emo{\vphantom{g}#1}}}
\newcommand{\emounbelief}[1]{{\emo{\vphantom{g}#1}}}
\newcommand{\emooutrage}[1]{{\emo{\vphantom{g}#1}}}
\newcommand{\emoguilt}[1]{{\emo{\vphantom{g}#1}}}

\usepackage{xspace}
\usepackage{amsmath}
\newcommand\ie{i.\,e.\xspace}
\newcommand\eg{e.\,g.\xspace}

\newcommand\cf{cf.\xspace}

\newcommand\US{U.\,S.\xspace}

	\usepackage[%capitalize,
                sort&compress]{cleveref}
    \crefname{chapter}{section}{sections}
	\Crefname{chapter}{Section}{Sections}
	\crefname{figure}{Figure}{Figures}
	\crefname{table}{Table}{Tables}

\newcommand{\norm}[1]{\left\lVert#1\right\rVert}

\usepackage{url}
\usepackage{balance} 
\usepackage{appendix}
\usepackage{booktabs}
\usepackage{stackengine}
\usepackage{tikz}
\newcommand*\numbersymbol[1]{\tikz[baseline=(char.base), font=\sffamily]{
            \node[fill=black, color=black, text=white, scale=0.75, shape=rectangle, draw, inner sep=2pt] (char) {#1};}}

\usepackage{placeins}

\title{Online Emotions During the Storming of the U.S. Capitol:\\ Evidence from the Social Media Network Parler}
\author{Anonymous Author(s)}

\author {
    Johannes Jakubik\textsuperscript{\rm 1},
    Michael Vössing\textsuperscript{\rm 1},
    Nicolas Pröllochs\textsuperscript{\rm 2},
    Dominik Bär\textsuperscript{\rm 3},
    Stefan Feuerriegel\textsuperscript{\rm 3}
    \\
}
\affiliations {
    \textsuperscript{\rm 1} Karlsruhe Institute of Technology \\
    \textsuperscript{\rm 2} University of Giessen \\
    \textsuperscript{\rm 3} LMU Munich \\
}

\begin{document}

\maketitle

\begin{abstract}
The storming of the U.S. Capitol on January 6, 2021 has led to the killing of 5 people and is widely regarded as an attack on democracy. The storming was largely coordinated through social media networks such as Parler. Yet little is known regarding how users interacted on Parler during the storming of the Capitol. In this work, we examine the emotion dynamics on Parler during the storming with regard to heterogeneity across time and users. For this, we segment the user base into different groups (e.g., Trump supporters and QAnon supporters). We use affective computing \cite{Kratzwald.2018} to infer the emotions in the contents, thereby allowing us to provide a comprehensive assessment of online emotions. Our evaluation is based on a large-scale dataset from Parler, comprising of 717,300 posts from 144,003 users. We find that the user base responded to the storming of the Capitol with an overall negative sentiment. Akin to this, Trump supporters also expressed a negative sentiment and high levels of unbelief. In contrast to that, QAnon supporters did not express a more negative sentiment during the storming. We further provide a cross-platform analysis and compare the emotion dynamics on Parler and Twitter. Our findings point at a comparatively less negative response to the incidents on Parler compared to Twitter accompanied by higher levels of disapproval and outrage. Our contribution to research is three-fold: (1)~We identify online emotions that were characteristic of the storming; (2)~we assess emotion dynamics across different user groups on Parler; (3)~we compare the emotion dynamics on Parler and Twitter. Thereby, our work offers important implications for actively managing online emotions to prevent similar incidents in the future. 
\end{abstract}

\section{Introduction}
On January 6, 2021, hundreds of violent rioters stormed the U.S. Capitol to overturn the defeat of Donald Trump in the 2020 presidential election \cite{Frenkel.06.01.2021}. Throughout the event, four rioters and one police officer were killed \cite{Romero.12.01.2021}, because of which the event is widely viewed as an unprecedented ``attack on democracy'' \cite{BBC.2021}.  

The storming of the U.S. Capitol was extensively discussed on social media, in particular, on Parler (\url{https://parler.com}). Parler is a self-proclaimed ``free speech'' social media platform where, analogously to other microblogging platforms, users can connect, like, and publicly exchange posts. However, Parler belongs to the group of alt-tech platforms, meaning that it presents itself as an alternative to mainstream social media networks such as Twitter \cite{Freelon.2020}. Owing to this, Parler has attracted a user base that consists primarily of conservatives, conspiracy theorists, and far-right extremists \cite{Romero.12.01.2021,Baer.2022}. Parler is regarded as the primary communication channel used during the storming of the U.S. Capitol \cite{Frenkel.06.01.2021}. Unsurprisingly, many posts from Parler contain highly emotional language (\eg, ``\texttt{God is with Us. READY TO SPILL MY BLOOD FOR AMERICA}'' or ``\texttt{WE ARE FORCEFULLY TAKING BACK AMERICA!}'').

As we hypothesize below, online emotions may have played an important role in the social media network Parler during the storming. This is stipulated by research from sociology and political science, suggesting that emotions play a central role in social movements and protests \citep{Jasper.2018,VanTroost.2013}, whereby people can experience emotions on behalf of their group \citep{VanTroost.2013}. For example, negative emotions (\eg, outrage) are often triggered by interactions of protesters with opponents, while positive emotions (\eg, joy) can result from the interaction with other in-movement activists. Notably, emotions can be elicited even if the individuals themselves are not directly confronted with the triggering situation \citep{VanTroost.2013}. Given the crucial roles of emotions in triggering protests, we investigate the emotion dynamics on Parler during the storming of the U.S. Capitol.

\textbf{Research questions:} 
In line with prior literature, we expect Parler's content to be characterized by specific emotions inherent to social movements and protests \citep{Goodwin.2009,Jasper.2018,VanTroost.2013}. We further hypothesize that there was considerable heterogeneity in online emotions across time and between specific subgroups of Parler users. For this purpose, we compare emotion dynamics of particularly vocal subgroups on Parler, such as Trump supporters, QAnon supporters, and supporters of alleged election fraud. We expect that these subgroups responded differently to the events on January 6, 2021, due to different attitudes and goals \cite{Papasavva.2021,NewYorkTimes.04.03.2021}. For example, some users may want to express their disappointment with election results but do not necessarily invoke riots to fight against democratic institutions. Similarly, certain user groups may support peaceful protests but are reluctant to participate in protests that turn violent. For this reason, different perceptions of the events unfolding during the storming of the \US Capitol should lead to different dynamics of online emotions.

We expect the overall Parler network to disagree with the storming and, therefore, shift toward negative emotions. In contrast, we expect extremist users to show negative emotions before the storming (as users express their disagreement with the political course). In response to the storming, we expect them to show more positive emotions (reflecting their approval) and again more negative afterward (expressing unbelief or outrage concerning the outcome). As a specific example, rioters often experience an emotional state of guilt after protests \citep{VanTroost.2013}. Hence, it is important to see whether users in the Parler network also use language characterized by guilt, or whether this emotion is absent. Motivated by this, we study the following research questions: 
\begin{quote}\begin{itemize}
\item \textbf{RQ1:} \emph{How did sentiment and online emotions in the social media network Parler change during the storming of the U.S. Capitol?}
\item \textbf{RQ2:} \emph{How did sentiment and online emotions vary across user groups?}
\item \textbf{RQ3:} \emph{How did emotion dynamics differ between Parler and Twitter?}
\end{itemize}\end{quote}

\textbf{Data:}\footnote{Sentiment data is available via \url{https://github.com/jhnnsjkbk/EmotionDynamics}}  We build upon a large dataset \cite{Aliapoulios.2021} from the Parler network. Our analysis focuses on all content created on January 6, 2021. The data contain 717,300 messages (so-called Parleys) from 144,003 different users. Our data also includes users' biographies, which we use to infer self-disclosed partisanship (\eg, whether users self-disclose to be Trump or QAnon supporters). The data collection and the analysis follow common standards for ethical research \cite{Rivers.2014}.

\textbf{Methods:}\footnote{Codes for reproducibility are available via \url{https://github.com/jhnnsjkbk/EmotionDynamics}.} In this work, we examine emotion dynamics on Parler during the storming of the U.S. Capitol. Specifically, we use affective computing to infer the emotions in Parleys. For each Parley, the degree of emotion is rated according to so-called basic emotions. Basic emotions refer to a subset of emotions that are universally recognized across cultures \cite{Ekman.1992,Sauter.2010} from which more complex emotions can be derived. We adopt Plutchik's wheel of emotions \cite{Plutchik.2001}, comprising of 8 basic emotions (\emoanticipation{anticipation}, \emosurprise{surprise}, \emoanger{anger}, \emofear{fear}, \emotrust{trust}, \emodisgust{disgust}, \emojoy{joy}, \emosadness{sadness}). We further study four specific derived emotions: \emodisapproval{disapproval}, \emounbelief{unbelief}, \emooutrage{outrage}, and \emoguilt{guilt}. This choice is based on prior research on emotions in protests as the selected emotions are particularly characteristic of protests. For example, \citet{Jasper.1998} identifies \emooutrage{outrage} as a reactive emotion in protests, triggered by activists toward concrete policies and decision-makers. In line with this, \citet{VanTroost.2013} identify \emoguilt{guilt} as an important emotion after protests, which can yield an individual to not transgress social norms and values in the future. We measure heterogeneity in emotions across time, across different user groups, and in comparison to Twitter to answer the above research questions.

\section{Related Work}
\label{sec:related_work}

\subsection{Online Social Media}

Online social media has evolved into a widespread source of information and opinion formation. However, even on popular platforms (\eg, Twitter), the majority of users regard the news to be largely inaccurate \cite{Mitchell.2020,Schaeffer.2020}. For instance, over 80\,\% of Republicans believe it is likely that platforms censor certain political viewpoints \cite{Schaeffer.2020}. Hence, the importance and downsides of social media for political discourse have been subject to extensive discussion \cite{Lazer.2018}.

Research has also studied social media use around specific events. Several studies have focused on political discourse on social media, especially before elections \citep[\eg][]{Bakshy.2015,Grinberg.2019,Recuero.2020}. These works analyze the exposure of users to misinformation and how the discourse is manifested in offline outcomes (\ie, votes). Other works focus on the public discourse around riots and social movements on social media, as well as the role of social media in these. Specific examples are, e.g., the Boston Marathon bombing \citep{Starbird.2017}, mass shootings \cite{McIlroy.2019,Starbird.2017}, protests and conflicts in Brazil \citep{Costa.2015}, and the Arab Spring \citep{Venkatesan.2021}.

Social media is characterized by a high degree of (affective) polarization \cite{MichaelConover.2011}, which is reflected in large parts of the American society \cite{Druckman.2019,Iyengar.2019}. One study found that interactions with political content reveal dynamics of strong polarization through a high degree of clustering within ideological groups \cite{MichaelConover.2011}. 

\subsection{Dynamics of Online Emotions}

Emotions, defined as the response to environmental stimuli \cite{Zhang.2013}, have been found to have vast influence over online behavior. This has been previously confirmed, \eg, for retweet dynamics \cite{Kim.2012,Stieglitz.2013} and online consumption \cite{Upworthy.2022}. Psychological research \cite{Lutz.2020} suggests that social media users rely primarily on affective information processing (rather than cognitive processing). As such, emotions are important drivers of online behavior \cite{Brady.2017,Naumzik.2022,ScientificReports.2021,EPJ.2021,Upworthy.2022,Solovev.2022,Zollo.2015}. For example, outrage is regarded as an especially important emotion for online behavior but with severe negative risks \citep{Crockett.2017}. Yet, to the best of our knowledge, the role of emotions on Parler and, in particular, during the storming of the U.S. Capitol has not yet been studied. Motivated by previous studies, we expect that the storming of the U.S. Capitol is characterized by unique emotion dynamics.

\subsection{Parler as an Alternative Social Media Platform}

Alternative social media platforms (``alt-tech'') provide a parallel space for ideas and beliefs outside the boundaries of speech permitted on mainstream social media platforms. Individuals can consume information aligned with their preexisting belief systems. For example, a large share of Parler users believes that the 2020 U.S. presidential election was fraudulent. Restrictions of content are one of the reasons that led to the popularity of alt-tech platforms (\ie, de-platforming) \cite{Pressman.29.06.2020}.  When Facebook and other platforms banned content related to the far-right QAnon conspiracy theory, a large number of proponents likely migrated to Parler \cite{Pressman.29.06.2020}. As a result, alt-right social media is characterized by increased user segregation, polarization, and hyperpartisanship \citep[\cf][]{Finkel.2020}.

Parler was extensively used before and during the storming \cite{Munn.2021,Frenkel.06.01.2021}. For example, the Daily Mail has generated a location map of Parler users during the storming \cite{Jewers.2021}. The analysis of GPS coordinates of posted videos showed that users made several posts from inside the Capitol. \citet{ProPublica.2021} conducted a detailed analysis, in which content was manually inspected, finding repeated documentation of violence. Other works provide an explanatory analysis of user engagement on Parler, namely the number of new user accounts \cite{Hitkul.2021}, the volume of posts \cite{Aliapoulios.2021,Otala.2021}, and user characteristics \cite{Baer.2022}. Motivated by the widespread use of Parler during the storming, we seek to understand the role of online emotions.

\section{Data}
\label{sec:data}

\subsection{Timeline of the Storming of the U.S. Capitol}\label{sec:timeline}

Table~\ref{tab:timeline} summarizes important events during the storming of the U.S. Capitol. Throughout our analysis, we refer to the start of the storming as 11:00 EST, \ie, the time when Trump supporters first headed to the U.S. Capitol building. We refer to the end of the storming as 19:30 EST, \ie, the time when the U.S. Capitol was declared secure. Based on this, we define the following three time periods: \underline{before} (up until 11:00 EST), \underline{during} (11:00 EST to 19:30 EST), and \underline{after} (from 19:30 EST) the storming. All timings in this paper are reported in Eastern Standard Time (EST) and have been previously converted.

\begin{table}[ht]
\centering
\footnotesize
\begin{tabular}{c ll}
  \toprule
    \textbf{Event} & \textbf{Time} & \textbf{Description} \\ 
  \midrule
    \numbersymbol{1} & 11:00 & Protesters head to the U.S. Capitol \\[0.3em] 
    \numbersymbol{2} & 12:00 & Speech of Donald Trump \\[0.3em] 
    \numbersymbol{3} & 14:15 & Rioters enter building of U.S. Capitol \\[0.3em]  
    \numbersymbol{4} & 16:17 & Donald Trump releases a video to mob \\
      &                     & (``Go home, we love you, [...]'')  \\[0.3em]  
    \numbersymbol{5} & 19:00 & Twitter and Facebook remove posts, and \\
      &                     & Twitter blocks Donald Trump's account \\[0.3em]  
    \numbersymbol{6} & 19:30 & U.S. Capitol declared secure \\[0.3em] 
    \numbersymbol{7} & 20:30 & Facebook blocks Donald Trump's account \\ 
   \bottomrule
   \multicolumn{3}{l}{\footnotesize  Time in Eastern Standard Time (EST)}
\end{tabular}
\caption{Timeline of key events during the storming of the U.S. Capitol.} 
\label{tab:timeline}
\end{table}

\subsection{Data Collection}

In this study, a large public dataset \cite{Aliapoulios.2021} from Parler is analyzed. We restrict our analysis to all content (\ie, Parleys) from January 6, 2021, the day of the storming. Overall, this corresponds to 717,300 Parleys from 144,003 users. In the robustness checks, we later examine other time periods. Our data comprises of the content of all Parleys, the corresponding timestamps, as well as the user biographies. 

\subsection{Segmentation of User Base}

We expect significant differences in the emotion dynamics between specific subgroups on Parler. The rationale is that, for example, Trump supporters, QAnon supporters, and supporters of alleged election fraud likely evaluated the events on January 6, 2021 differently due to varying attitudes and goals \cite{Papasavva.2021,NewYorkTimes.04.03.2021}. For instance, some user groups might endorse peaceful protests but have concerns when those protests turn violent. Likewise, users aiming to express their disappointment with election outcomes do not necessarily approve attempts to overthrow democratic institutions. Overall, different appraisals of the unfolding events during the storming should result in different emotional reactions across time and user groups.

We therefore subset user groups on Parler that are particularly relevant in the context of the storming of the U.S. Capitol \cite{Wells.08.01.2021}. Specifically, we follow \citet{Aliapoulios.2021}and \citet{Ng.2021} by basing our analysis on the following user groups: (i)~Trump supporters, (ii)~QAnon supporters, and (iii)~supporters of alleged election fraud. In order to assign individual users to the corresponding groups, we search for particular hashtags in a user's biography. We select users based on hashtags in their biography instead of, for example, in their posts, as this makes sure that the users \emph{self-identify} themselves with the hashtags. Analogously to prior research \citet{Aliapoulios.2021,Baer.2022,Ng.2021}, we use the following hashtags to identify users belonging to the different user groups 
(see Table~\ref{tab:summary-usergroup} for summary statistics):

\begin{itemize}
\item \textbf{Trump supporters} (hashtags: \#Trump2020, \#maga, \#trump) represent a large share of the user base, which is in line with the earlier discussion stating that Parler has a significant conservative user base \cite{Romero.12.01.2021}.
\item \textbf{QAnon supporters} (hashtags: \#thegreatawakening, \#wwg1wga, \#qanon) believe in a far-right conspiracy theory which, among others, postulates that a ``shadowy cabal of Democratic pedophiles'' plotted against Trump while he was in office \cite{NewYorkTimes.04.03.2021}.
\item \textbf{Alleged election fraud} supporters (hashtags: \#stopthesteal, \#voterfraud, \#electionfraud) believe in widespread fraud during the 2020 presidential election \cite{Frenkel.06.01.2021}.
\end{itemize}

\noindent
Of note, the user groups are largely disjunct. For instance, there is only a small overlap between Trump supporters and supporters of alleged election fraud (\ie, 9\,\% of users overlap). The overlap between Trump supporters and QAnon supporters amounts to 12\,\%. Only 4\,\% of the users overlap between QAnon supporters and supporters of alleged election fraud. In the robustness checks, we repeat our analysis where users with overlap are removed, yielding consistent findings.

\begin{table}[ht]
\centering
\scriptsize
\begin{tabular}{lrrr}
  \toprule
\textbf{User group} & \textbf{\#Users (Share)} & \textbf{\#Posts} & \textbf{\#Posts/user} \\ 
  \midrule
  All (= overall Parler network) & 144,003 (100\%) & 717,300 & 4.98 \\[2pt]
  \midrule
  Trump supporters & 972 (0.7\%) & 6,794 & 6.99 \\
  Alleged election fraud  & 316 (0.2\%) &  2,219  & 7.02 \\
  QAnon supporters & 391 (0.3\%) & 2,642  & 6.76 \\
  
  \bottomrule
\end{tabular}
\caption{Segmentation of user base (for all users with at least one post on January 6, 2021).} 
\label{tab:summary-usergroup}
\end{table}

\subsection{Descriptives}

\begin{table*}[ht]
\scriptsize
\centering
\begin{tabular}{lp{0.8\textwidth}l}
\toprule
\textbf{Time} & \textbf{Content} & \textbf{\#Reactions} \\ \midrule \rule{0pt}{0ex}
06:56        & ``\texttt{Mike Pence is a TRAITOR \& must be charged with TREASON this morning. We The People demand it TODAY.}'' & 602               \\ \rule{0pt}{3ex}

12:59        & ``\texttt{TRUMP in DC: ‘We Will Never Give Up, We Will Never Concede, Our Country Has Had Enough’}'' & 947               \\ \rule{0pt}{3ex}

14:24        & ``\texttt{The time has come Patriots. This is our time. Time to take back our country. Time to fight for our freedom. Pledge your lives, your fortunes, \& your sacred honor. There will not be another chance. Speak TRUTH. Be FEARLESS. Almighty God is with you. TODAY IS OUR DAY.}'' & 1274              \\ \rule{0pt}{3ex}

16:08        & ``\texttt{Those who breached the Capitol Building and committed acts of violence have done grave damage in many ways. This will all need to be sorted out, the violent from the peaceful, but the violent must be punished. And the media must be careful not to paint everyone with the same broad brush. But make no mistake, the perpetrators must be punished.}''   & 3033  \\ 
\bottomrule 
\multicolumn{3}{l}{\scriptsize Time in Eastern Standard Time (EST); The reaction count refers to the number of reposts and comments the post received in the first hour after it was created.
}
\end{tabular}
\caption{Example messages (Parleys) from the day of the storming of the U.S. Capitol. 
} 
\label{tbl:posts_examples}
\end{table*}

The crucial role of Parler during the storming of the U.S. Capitol is anecdotally illustrated through exemplary Parler messages in Table~\ref{tbl:posts_examples}. Overall, the examples highlight the presence of affective, highly polarized language. We observe that users frequently use partisan language (\eg, ``\texttt{Patriots}'', ``\texttt{Truth}'') that is aggressive or even violent (\eg, ``\texttt{READY TO SPILL MY BLOOD FOR AMERICA}''). 

\begin{figure}[ht]
    \centering
    \includegraphics[trim = 0mm 50mm 0mm 60mm, clip, width=\columnwidth]{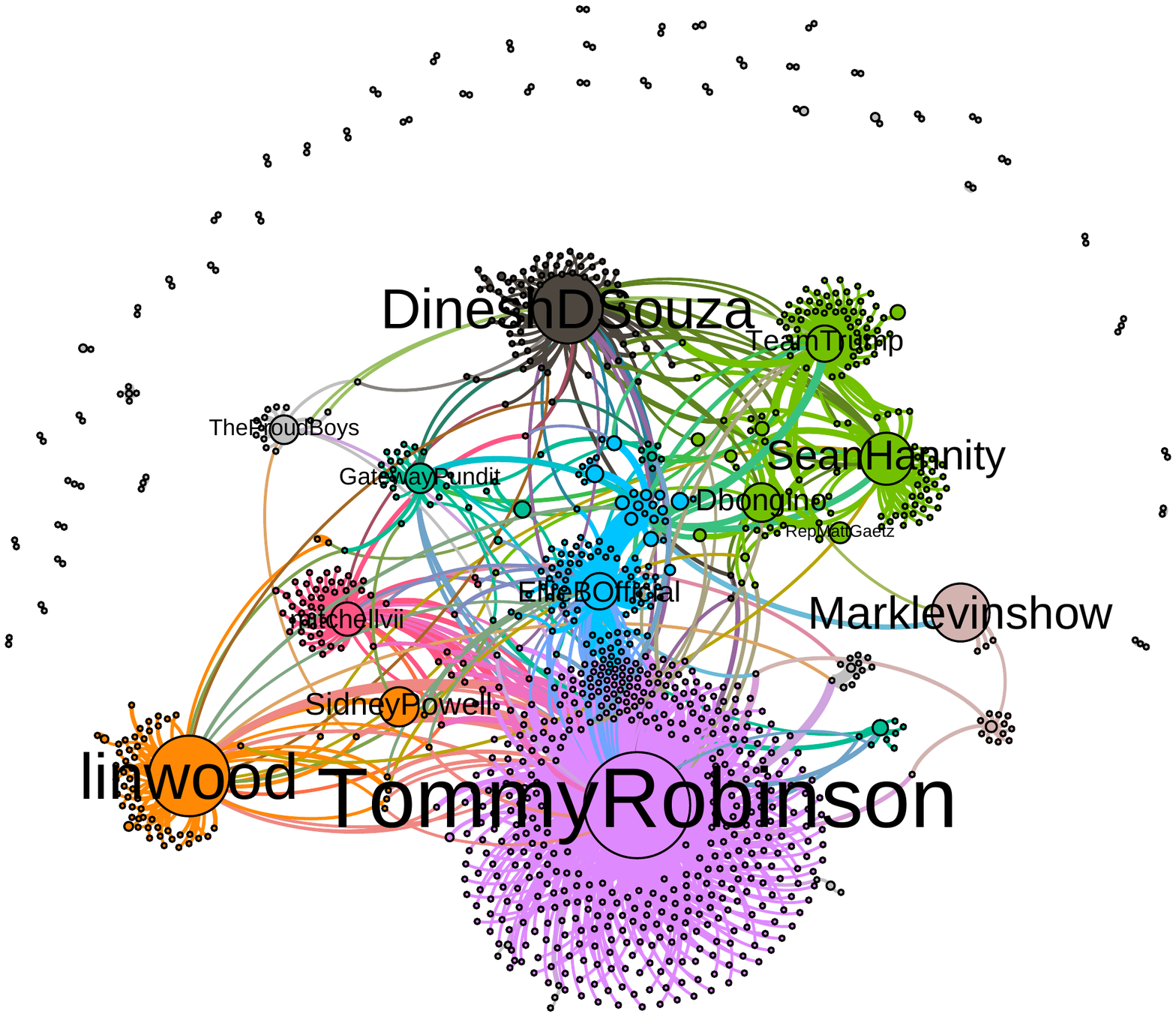}
    
    \begin{tabular}{p{\columnwidth}} 
    {\scriptsize \emph{Notes:} The size of the nodes represents the number of posts created by the respective users. The edges depict interactions (\ie, re-posts or comments) between users. For readability, only users who interacted at least 10 times with another user are shown. The nodes were placed using the ForceAtlas2 algorithm \cite{Jacomy.2014}. The colors indicate communities of users based on a modularity optimization heuristic. \par}
    \end{tabular}

    \caption{Social network plot showing post of users and their interactions (\ie, reposts, comments) on January~6, 2021.}
    \label{fig:network}
\end{figure}

The Parler network is illustrated in Fig.~\ref{fig:network}. The plot shows the volume of posts created by individual users as well as the volume of interactions between users (\ie, comments, reposts) on January 6. Evidently, posts were primarily created by a small number of influential users. For example, Tommy Robinson is a British far-right activist and former leader of the English Defence League (username: TommyRobinson; posts on January~6: 126; reactions on January~6: 47,593) and Bill Mitchell is a conservative pundit, who has been banned from Twitter for his controversial messages (username: mitchellvii; posts: 116; reactions: 11,720). 

\subsection{Ethical Considerations}

The dataset used in this study is public, open access, and de-identified (\ie, the names of all user accounts were removed). For details, we refer to \citet{Aliapoulios.2021}. The collection of the dataset was conducted following legal standards \cite{Aliapoulios.2021}, as well as standards for ethical research \cite{Rivers.2014}.  

We respect the privacy and agency of all people potentially impacted by this work and take specific steps to protect their privacy. We do not identify users that have chosen to remain anonymous by not verifying their account. We primarily analyze data and report results in aggregate. By accepting Parler's privacy policy from Parler, users agree that their content is ``public by default''.\footnote{Parler Privacy Policy: \url{https://legal.parler.com/documents/privacypolicy.pdf}, accessed October 21, 2021.} Moreover, Parler cautions users to ``carefully consider whether and what to post and how you identify yourself on our Services.'' The analysis was conducted in accordance with the Institutional Review Board at ETH Zurich.

We further acknowledge that capturing emotion dynamics could be used for surveillance or over-policing of non-harmful movements in social media. To mitigate potential risks, we analyze emotion dynamics at the aggregate level and not at the individual level. We further advise users of social media to carefully consider what and under what name they disclose before posting, thus taking into account potential benefits as well as risks. Research and practice need to analyze data carefully (e.g., focus on aggregate analyses), so that the benefits of the analyses outweigh potential risks.

\section{Methods}
\label{sec:methods}

\subsection{Emotion Model}
In this work, we use Plutchik's emotion model \cite{Plutchik.2001}, which has been widely adopted in psychology as well as affective computing \cite{Kratzwald.2018,Mohammad.2021,ScientificReports.2021,EPJ.2021,Upworthy.2022}. The model groups emotions into (1)~\textbf{sentiment} to distinguish positivity vs. negativity; (2)~\textbf{basic emotions}; and (3)~\textbf{derived emotions} inferred from the basic emotions. 

\textbf{Sentiment:} Plutchik's model distinguishes positive and negative emotions. When making such binary classification by overall valence, we refer to it as sentiment. 

\textbf{Basic emotions:} In emotion theory, basic emotions are a subset of emotions that are universally recognized across cultures. Plutchik's model defines 8 basic emotions. The basic emotions are \emoanger{anger}, \emoanticipation{anticipation}, \emodisgust{disgust}, \emofear{fear}, \emojoy{joy},  \emosadness{sadness}, \emosurprise{surprise}, and \emotrust{trust}.

\textbf{Derived emotions:} More complex emotions can be derived from the basic emotions \cite{Ekman.1992,Sauter.2010}. In Plutchik's model, there are 24 derived emotions. Each derived emotion combines two basic emotions. For instance, \emounbelief{unbelief} is a combination of \emodisgust{disgust} and \emosurprise{surprise}. 

In this work, we focus on four derived emotions: \emodisapproval{disapproval}, \emounbelief{unbelief}, \emooutrage{outrage}, and \emoguilt{guilt}. This choice was made based on prior research on emotions in social movements. Importantly, the selected emotions are particularly characteristic of protests and ought to capture the emotions from the rioters \citep[\cf][]{Flam.2007,Jasper.2018,VanTroost.2013}. For example, \citet{Jasper.1998} identifies \emooutrage{outrage} as a primarily reactive emotion in protests, triggered by activists who aim at transforming anxieties and fears into outrage toward concrete policies and decision-makers. Moreover, \emooutrage{outrage} is especially dominant in online environments \citep{Crockett.2017}. In a similar vein, \citet{VanTroost.2013} identify \emoguilt{guilt} as an important emotion after protests that can yield an individual, for example, to not transgress social norms and values in the future. The selected emotions are further relevant in psychology. The reason is that they ship negative emotions that are of high-arousal nature and, different from low-arousal emotions, are especially strong stimuli \citep{Kratzwald.2018}. As a result, they are effective for drawing attention, even in online spheres, and thus lead to behavioral responses among recipients \citep{EPJ.2021,Upworthy.2022}.

\subsection{Affective Computing}

We use affective computing \citep[\cf][]{Kratzwald.2018} to infer emotions in text. Specifically, we implemented a dictionary approach as in \citet{Mohammad.2021}. Dictionary approaches are used when the objective is to make explanatory (rather than predictive) inferences and where it is important to obtain reliable interpretations \citep[\eg][]{Brady.2017,ScientificReports.2021,EPJ.2021,Upworthy.2022,Vosoughi.2015}. In our work, the NRC emotion lexicon was used due to its comprehensiveness, granularity, and widespread use \cite{Mohammad.2021}. In particular, there is a shortage of alternative approaches (\eg, machine learning) for quantifying basic and derived emotions from Plutchik's emotion \cite{Mohammad.2021}.

We proceeded as follows. First, we tokenized the content from Parler and counted how frequently different terms from the dictionary occurred. This resulted in an 8-dimensional vector $n_i$ for Parler post $i$, where each element in the vector denotes the frequency of terms from one of the eight basic emotions. Afterward, we divided the word count by the total number of dictionary words in the text. This performed a normalization so that the values in the resulting vector sum to one across the basic emotions. Formally, $\rho_i = \frac{1}{\norm{n_i}_1}{n_i} \in [0,1]^8$ for a Parler post $i$. As a result, the values in $\rho_i$ range from zero to one and denote the intensity of the different basic emotions. For instance, a Parler post might thus contain 40\,\% \emoanger{anger} and 60\,\% \emofear{fear}. Altogether, this approach yields 8 scores measuring the relative intensity of basic emotions: $\mathit{Anger}_{i}$, $\mathit{Anticipation}_{i}$, $\mathit{Joy}_{i}$, $\mathit{Trust}_{i}$, $\mathit{Fear}_{i}$, $\mathit{Surprise}_{i}$, $\mathit{Sadness}_{i}$, and $\mathit{Disgust}_{i}$. 

Based on the above values, we then compute scores for (1)~\textbf{sentiment}, (2)~\textbf{basic emotions}, and (3)~\textbf{derived emotions} as follows:
\begin{enumerate}
\item \textbf{Sentiment.} According to Plutchik's emotion model \cite{Plutchik.2001}, basic emotions are either positive or negative. Based on this, we calculate an overall sentiment score that quantifies the overall valence of the content, that is, whether it conveys a positive or negative polarity. Formally, the sentiment for post $i$ is computed as the difference between positive and negative emotions:   
\begin{equation}
\mathit{Sentiment}_{i} =  \mathit{Positive}_{i} - \mathit{Negative}_{i} . 
\end{equation} 
\item \textbf{Basic emotions.} The basic emotions are simply the variables from above, namely $\mathit{Anger}_{i}$, $\mathit{Anticipation}_{i}$, $\mathit{Joy}_{i}$, $\mathit{Trust}_{i}$, $\mathit{Fear}_{i}$, $\mathit{Surprise}_{i}$, $\mathit{Sadness}_{i}$, and $\mathit{Disgust}_{i}$.
\item \textbf{Derived emotions.} 
In Plutchik's emotion model, the derived emotions are defined as \emph{combinations} of two basic emotions (so-called emotional dyads), because of which we compute the derived emotions as the sum of both basic emotions (\eg, $\mathit{Outrage}_{i} = \mathit{Anger}_{i} + \mathit{Surprise}_{i}$). Here, choosing the mean is supposed to capture the concept of a ``combined'' emotion and is analogous to earlier research \citep{EPJ.2021}. 

\end{enumerate}
Our analysis is based on relative emotional levels. Therefore, our results are independent of the volume of posts. This becomes particularly relevant as the volume of posts in social networks can significantly differ across time periods.

\subsection{Empirical Validation of Emotion Recognition}
\label{sec:validation_study}

Our results rely on the accuracy of our emotion recognition. We thus conduct a qualitative check using external labelers to verify how (i)~perceived sentiment, (ii)~perceived basic emotions, and (iii)~perceived derived emotions align with our above approach for affective computing. For this, we conducted three user studies using the online survey platform Prolific (\url{https://prolific.co}). For all studies, we randomly sampled 100 Parleys and presented them to $n = 7$ participants (English native speakers). The participants were asked to rate the perceived sentiment, basic emotions, or derived emotions in each Parley. Here, we employ a Likert scale from $-3$ to $+3$ ($-3$ indicates negative sentiment or absence of the specific emotion, while $+3$ refers to a positive sentiment or presence of the emotion).

For sentiment, the participants exhibited a statistically significant interrater agreement according to Kendall’s $W$ ($p < 0.01$). Furthermore, we compute the correlation between the human labels and the dictionary-based sentiment score. We found Spearman’s correlation coefficient to be positive ($r_s = 0.23$) and statistically significant ($p < 0.01$). For basic emotions, the participants exhibited a statistically significant interrater agreement according to Kendall’s $W$ for the aggregated basic emotions ($p < 0.05$). Overall, the correlation between the dictionary-based emotion scores and human annotations is $r_s = 0.09$ ($p < 0.05$) and thus statistically significant at common significance thresholds. For the derived emotions, the participants exhibited a statistically significant interrater agreement according to Kendall's $W$ ($p<0.01$). In addition, we found Spearman's correlation coefficient to be positive ($r_s = 0.05$) and statistically significant ($p < 0.05$) for the derived emotions. In sum, the relationships are statistically significant, implying that our dictionary approach overlaps with the emotions perceived by readers. 

\section{Results}
\label{sec:results}

\subsection{RQ1: How did sentiment and online emotions in the social media network Parler change during the storming of the U.S. Capitol?}

We now analyze how the average sentiment and emotions in Parleys changed during the storming.

\textbf{Sentiment:} 
In the following, we use Kolmogorov-Smirnov (KS) tests to study whether differences in sentiment before, during, and after the storming (see Table~\ref{tab:timeline}) are statistically significant. For the KS-tests, we consider the sentiment in each of these time periods as a specific probability distribution and compare their empirical distribution functions, and we test that the distributions are mutually independent, respectively.

\begin{figure}[ht]
    \centering
    \includegraphics[width=\columnwidth]{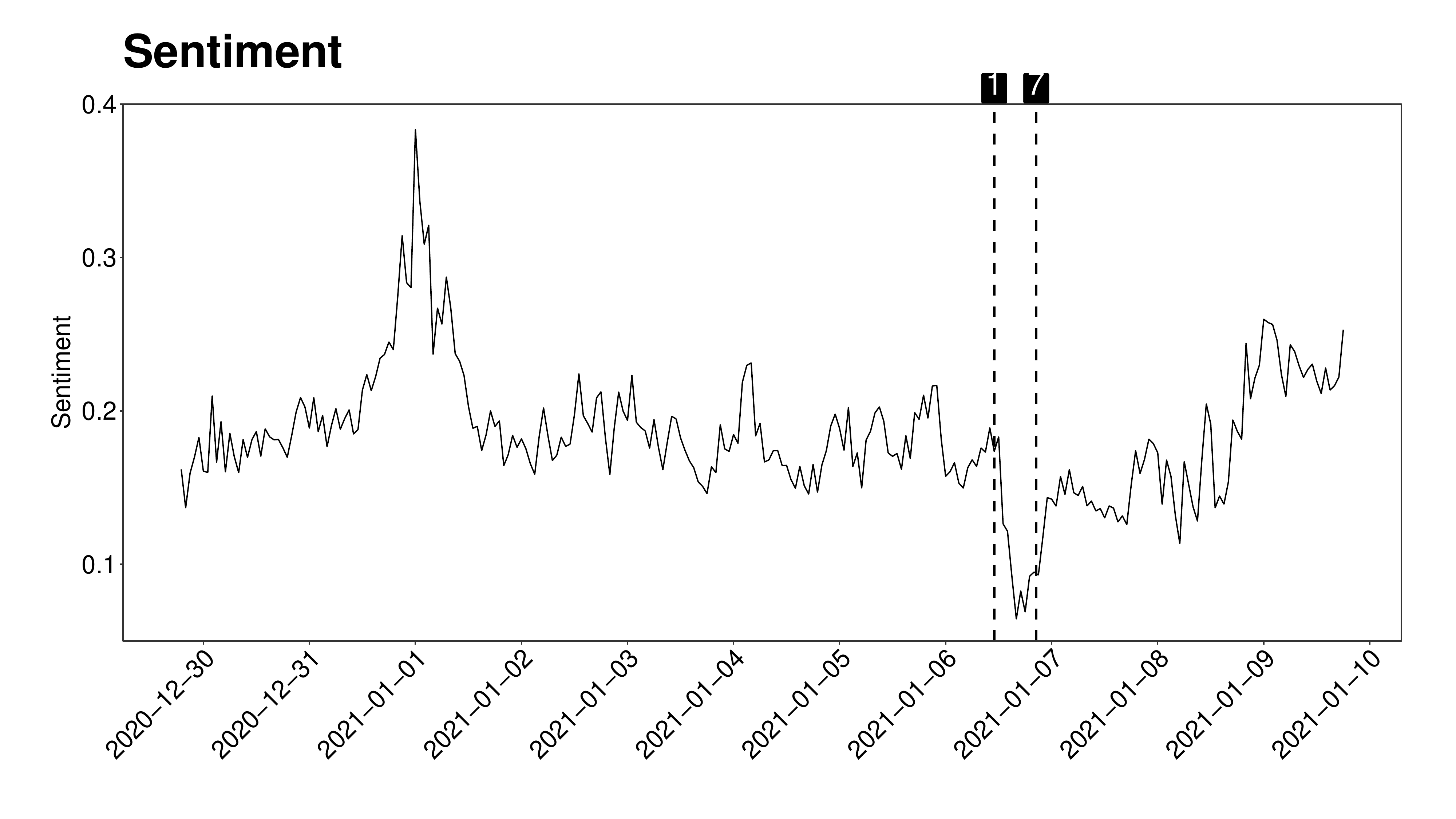}
    \caption{Sentiment (hourly) in Parler network between December 30, 2020 and January 9, 2021.}
    \label{fig:time_series_sentiment}
\end{figure}

Fig.~\ref{fig:time_series_sentiment} shows the sentiment on Parler between December 30, 2020 and January 9, 2021. On January 6, the day of the storming, there was a considerable decline in sentiment, which was pronounced when rioters were entering the Capitol (Event \numbersymbol{3}). This is confirmed via a KS-test {($D=0.183$, $p<0.001$)}. We observe a shift toward a more positive sentiment after the Capitol was declared secure (Event \numbersymbol{6}), as well as after Trump was banned from other platforms (Event \numbersymbol{5} and \numbersymbol{7}) {(KS-test: $D=0.169$, $p<0.001$)}. Eventually, it returned to the same levels as before the storming. The plot shows a particularly positive sentiment around New Year's Eve. This can be expected due to the surrounding celebrations and thus adds to the validity of our results. 

\begin{figure}[ht]
\captionsetup{position=top}
\centering
	\subfloat[Basic emotions]{
\begin{tabular}[c]{@{}l@{}}
\includegraphics[width=\columnwidth]{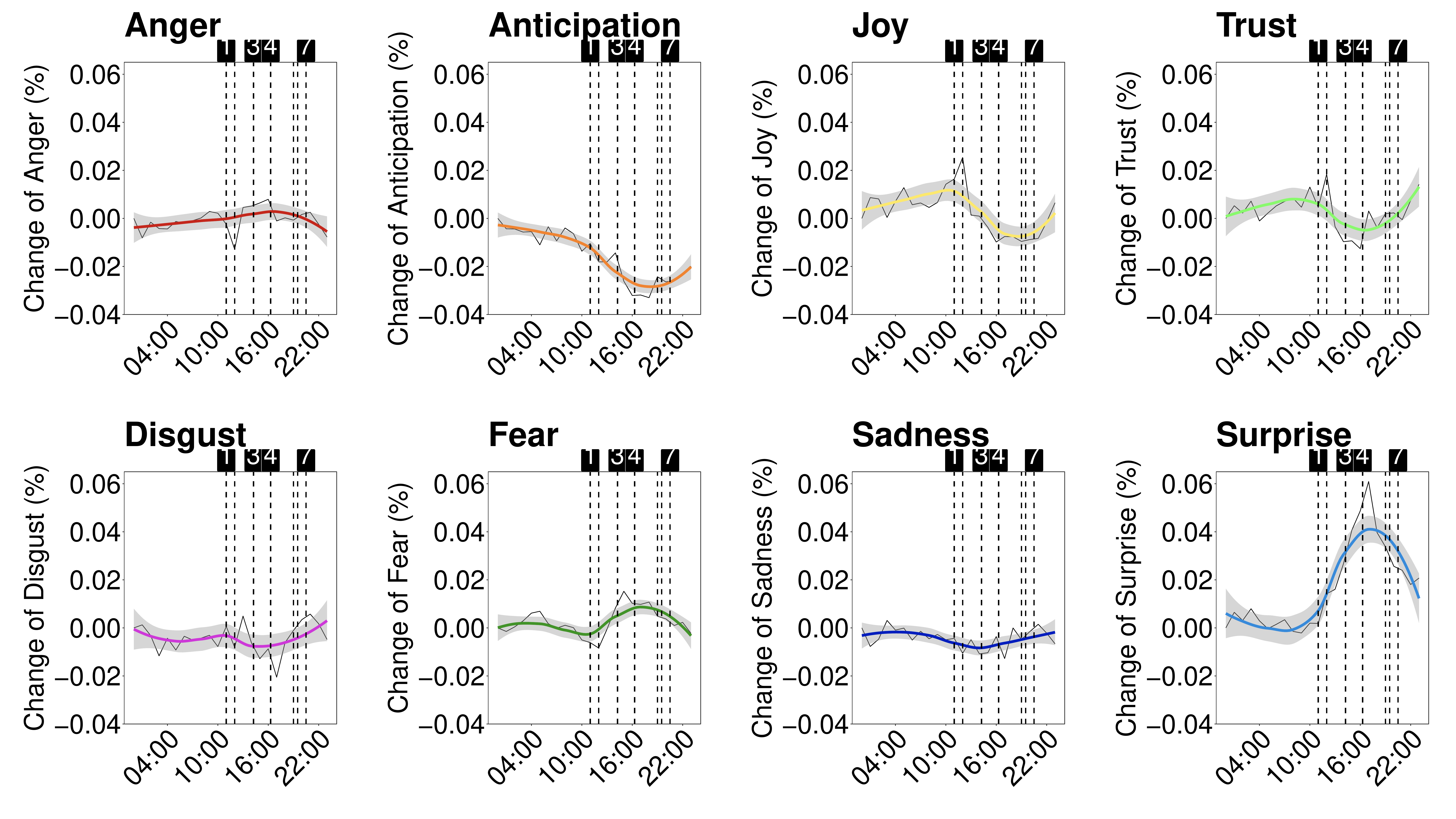}\\%
\end{tabular}%
} \hfill
\subfloat[Derived emotions]{
\begin{tabular}[c]{@{}l@{}}
%\\
\includegraphics[width=\columnwidth]{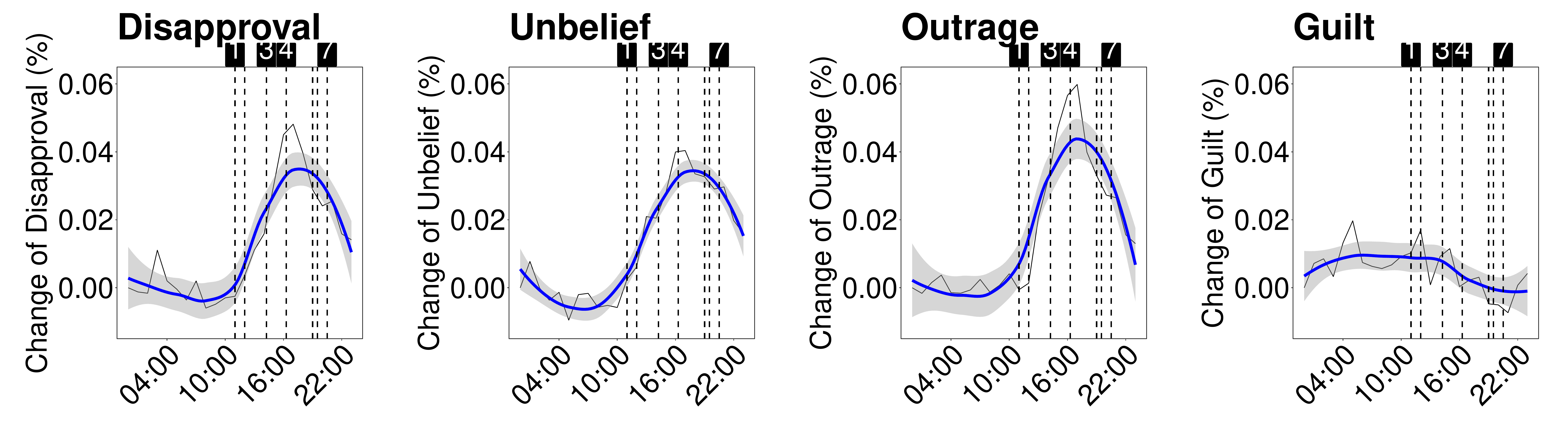}\\%
\end{tabular}%
} \hfill
\caption{Dynamics of (a) basic emotions and (b) derived emotions in the Parler network on January 6, 2021. Shown is the relative emotion level (in black). Shown (in color) is also a locally weighted scatterplot smoothing (LOESS) \cite{Jacoby.2000}.
}
\label{fig:time_series_emotions_change}
\end{figure}

\textbf{Basic emotions:} We now provide a granular assessment of emotion dynamics across the 8 different {basic emotions} (Fig.~\ref{fig:time_series_emotions_change}~(a)). For brevity, we show only the day of the storming (but note that we observe again peaks toward positive emotions around New Year's Eve). The level of \emojoy{joy} rose in anticipation of Donald Trump's speech (Event \numbersymbol{2}). Afterward, when rioters entered the building (Event \numbersymbol{3}), the level of \emofear{fear} increased steadily. The level of \emosurprise{surprise} peaked when Donald Trump released a message to his followers, postulating that they should ``\texttt{Go home, [...]}'' (Event \numbersymbol{4}). The observations are confirmed by KS-tests with $p<0.001$.

\textbf{Derived emotions:} 
In the following, we elaborate on four derived emotions: \emodisapproval{disapproval}, \emounbelief{unbelief}, \emooutrage{outrage}, and \emoguilt{guilt}. We chose these analogous to prior research as they are particularly characteristic of social movements and ought to capture the distinctive emotion dynamics of the rioters \citep{Jasper.2018,VanTroost.2013}. After the first protesters entered the Capitol building (Event \numbersymbol{3}), we find that the content conveys significantly more emotional language for \emodisapproval{disapproval} {(KS-test: $D=0.176$, $p<0.001$)}, \emounbelief{unbelief} {(KS-test: $D=0.221$, $p<0.001$)}, and \emooutrage{outrage} {(KS-test: $D=0.203$, $p<0.001$)}. Conversely, the level of \emoguilt{guilt} remained relatively constant throughout the storming. This again is confirmed by KS-tests.

\subsection{RQ2: How did sentiment and online emotions vary across user groups?}

\begin{figure}[ht]
\captionsetup{position=top}
\centering
	\subfloat[Sentiment]{\label{fig:heatmap_sentiment}
\begin{tabular}[c]{@{}l@{}}
\includegraphics[width=\columnwidth]{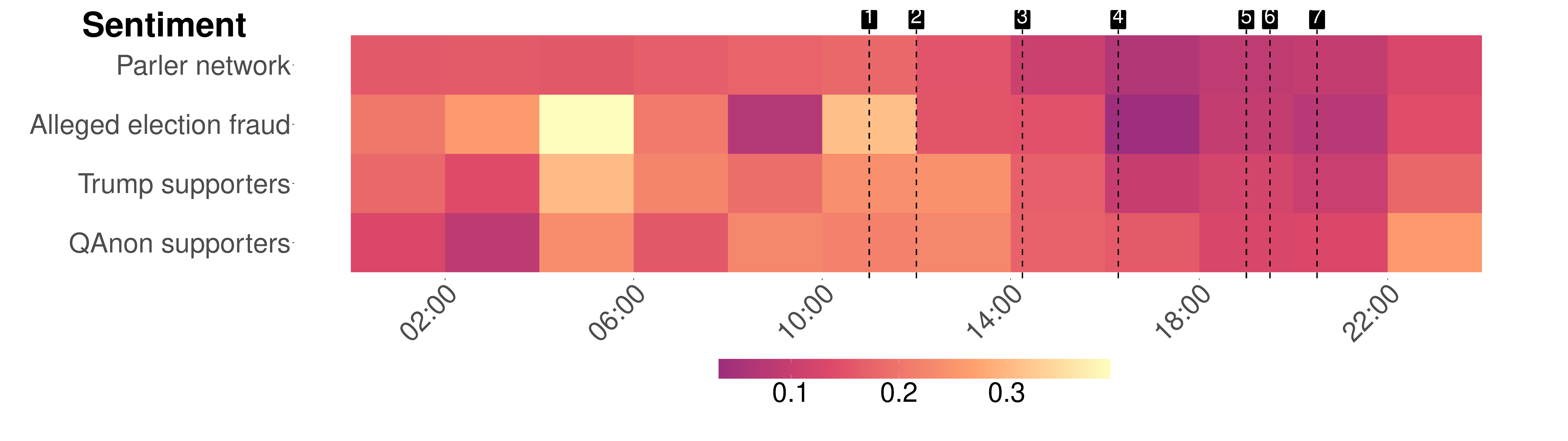}\\%
\end{tabular}%
} \hfill
\subfloat[Derived emotions]{\label{fig:heatmaps_derived_emotions}
\begin{tabular}[c]{@{}l@{}}
\includegraphics[width=\columnwidth]{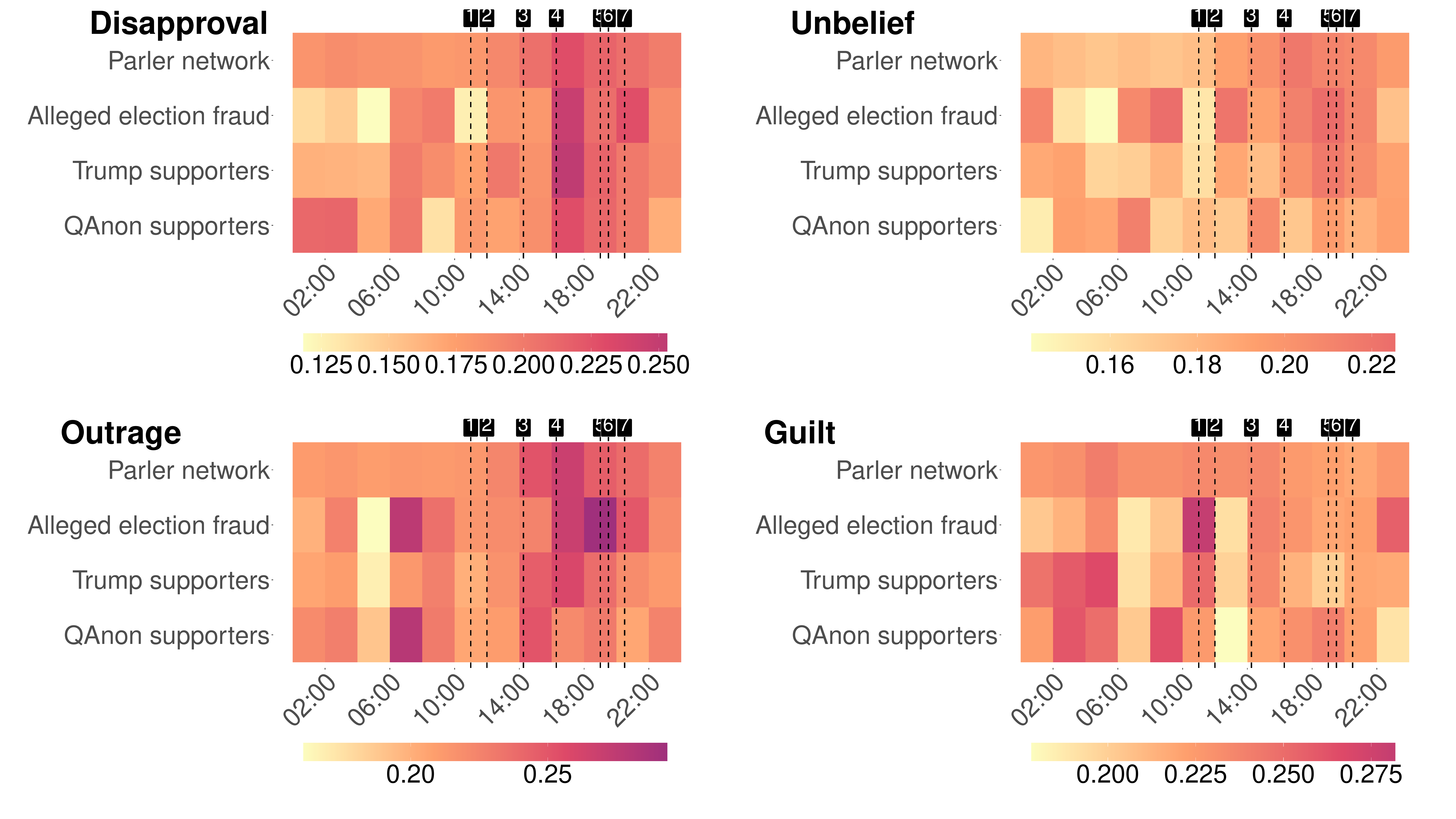}\\%
\end{tabular}%
} \hfill
\caption{Emotion dynamics (hourly) of (a) sentiment and (b) derived emotions for different user groups.
}
\label{fig:heatmaps_grouped}
\end{figure}

Fig.~\ref{fig:heatmaps_grouped} compares the sentiment and emotion dynamics across time and user bases. 

\textbf{Sentiment:} 
We find similar patterns for the sentiment of Trump supporters and supporters of alleged election fraud; see Fig.~\ref{fig:heatmaps_grouped}(a). The sentiment decreases throughout the storming and then remains at a comparatively negative level. For all user groups except QAnon supporters, we observe the most negative level after rioters entered the Capitol building (Event \numbersymbol{3} and \numbersymbol{4}). In contrast, the sentiment of QAnon supporters is more negative after the Capitol building is declared to be secure (Event \numbersymbol{6}). Hence, the user groups showed different emotional responses to the storming. 

For the Parler network as a whole, the sentiment was more negative during and after the storming of the Capitol. The number of words in posts during the storming that embed a negative emotion increased by 72.67\,\%. KS-tests confirm that the differences in the distributions of sentiment are statistically significant; both when comparing the time periods before vs. during the storming {(KS-test: $D=0.052$, $p<0.001$)} and before vs. after the storming {(KS-test: $D=0.059$, $p<0.001$)}. We find similar patterns for Trump supporters and supporters of alleged election fraud, for whom the sentiment was also more positive before the storming (compared to during or after the storming). The number of negative words in posts during the storming increased by 64.09\,\% for Trump supporters and by 38.26\,\% for supporters of alleged election fraud. For both groups, KS-tests confirm that the differences are statistically significant. In contrast, we observe no statistically significant difference for QAnon supporters. Trump supporters and QAnon supporters thus reveal different sentiment dynamics.

\textbf{Basic emotions:} 
The results for basic emotions are omitted for reasons of space. In contrast to the overall network, we observe, for instance, high levels of \emofear{fear} in Parleys that have been posted by supporters of alleged election fraud \emph{after} the storming of the Capitol. KS-tests confirm that the differences in the distributions before vs. after the storming are statistically significant {(KS-test: $D=0.112$, $p=0.006$)}.

\textbf{Derived emotions:}
Fig.~\ref{fig:heatmaps_grouped}(b) compares the emotion dynamics for derived emotions. We observe the highest levels of \emounbelief{unbelief} and \emooutrage{outrage} in posts of supporters of the alleged election fraud when the Capitol building was declared secure (Event \numbersymbol{6}). This is different for Trump supporters. The level of \emooutrage{outrage} is especially pronounced after rioters entered the Capitol building (Event \numbersymbol{3} and \numbersymbol{4}). QAnon supporters are an exception as their Parleys embed shallow levels of \emounbelief{unbelief} and \emooutrage{outrage} before the Capitol building is declared secure.

We again analyze whether the differences in sentiment before, during, and after the storming are statistically significant. For the overall Parler network, the emotional states \emph{during} the storming were characterized by \emodisapproval{disapproval}, \emooutrage{outrage}, \emoguilt{guilt}, and \emounbelief{unbelief}. For each of these emotions, KS-tests confirm that the differences in the distributions before vs. during the storming and before vs. after the storming are statistically significant. We found the opposite for supporters of alleged election fraud. Here, the levels of \emodisapproval{disapproval} and \emoguilt{guilt} were larger \emph{after} the storming of the Capitol. The emotional dynamic of Trump supporters resembles those of the overall Parler network in terms of \emodisapproval{disapproval}. However, in contrast to the overall network, Trump supporters embed a larger level of \emounbelief{unbelief} in their posts \emph{after} the storming than before and during the storming. Again, KS-tests confirm that these differences are statistically significant. QAnon supporters show a different pattern compared to the overall network. They embedded low levels of \emodisapproval{disapproval}, \emounbelief{unbelief}, and \emooutrage{outrage} \emph{after} the storming of the Capitol. As with sentiment, we observe no statistically significant differences in the emotion levels before, during, and after the storming for QAnon supporters.

\subsection{RQ3: How did emotion dynamics differ between Parler and Twitter?}

\begin{figure}[ht]
\captionsetup{position=top}
\centering
	\subfloat[Sentiment]{\label{fig:time_series_tweet_sentiment}
\begin{tabular}[c]{@{}l@{}}
    \includegraphics[width=0.8\linewidth]{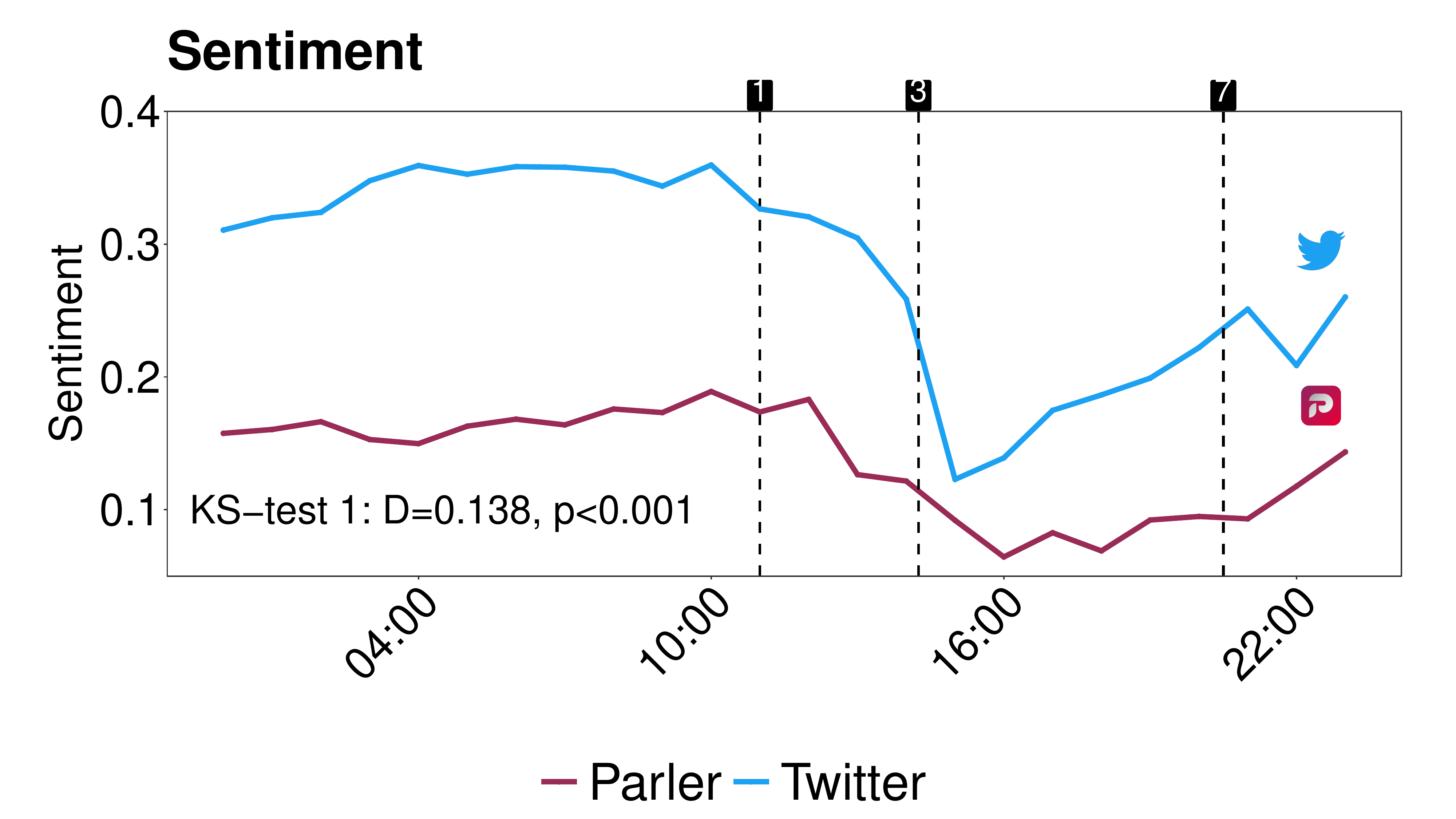}\\
\end{tabular}%
} \hfill
\subfloat[Derived emotions]{\label{fig:time_series_tweet_derived}
\begin{tabular}[c]{@{}l@{}}
    \includegraphics[width=\linewidth]{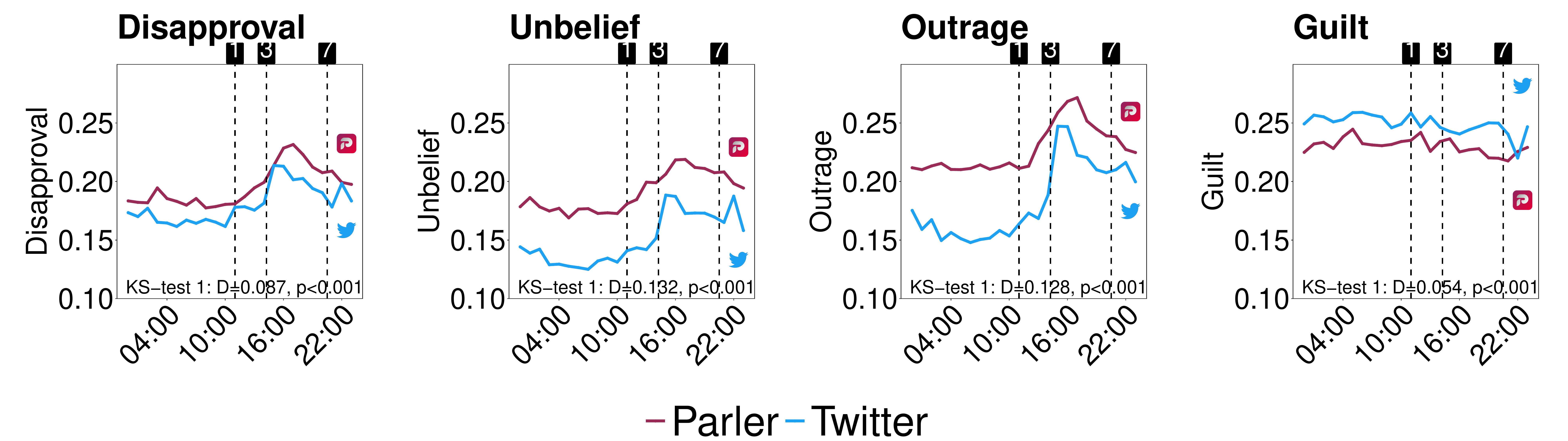}\\
\end{tabular}%
} \hfill
\caption{
Emotion dynamics (hourly) of (a) sentiment and (b) derived emotions on Parler and Twitter on January~6, 2021.
}
\label{fig:emotions_crossplatform}
\end{figure}

Next, we perform a cross-platform analysis and compare the emotion dynamics on Parler and Twitter. For this purpose, we use the Twitter Historical API to collect a time series of $5,000$ random tweets in English from the \US for each hour on January 6, 2021. This process yielded $5,000 \times 24 = 120,000$ tweets. We then use the above approach to extract sentiment and emotion scores for each tweet analogous to the approach above. Recall that sentiment (as well as emotions) are computed as rates, and thus standardized by length, so that the different text lengths in Parler vs. Twitter posts allow for direct comparisons across platforms. The comparison of the emotion dynamics from Parler and Twitter is shown in Fig.~\ref{fig:emotions_crossplatform}. 

\textbf{Sentiment:} We observe statistically significant differences with regard to sentiment on January 6, 2021 between Parler and Twitter. In absolute terms, the sentiment on Parler was significantly more negative than on Twitter throughout the entire day (KS-test: $D=0.138$, $p<$0.001). While both user bases responded negatively to the unfolding events during the storming of the Capitol, the (negative) reaction of Parler users was less pronounced. Specifically, compared to Twitter, we observe a smaller difference in the average sentiment before and during the storming (KS-test: $D=0.625$; $p<0.1$). Interestingly, we also find that the (negative) response on Parler lagged behind the (negative) response on Twitter. The most negative sentiment on Twitter is observed after rioters entered the Capitol, whereas the most negative sentiment on Parler is observed shortly before Trump releases a video to the rioters (``\texttt{Go home, [...]}'').

\textbf{Basic emotions:} The results for basic emotions are omitted for reasons of space. In short, we find that emotions like \emoanger{anger}, \emodisgust{disgust} and \emofear{fear} are significantly more pronounced on Parler compared to Twitter, while positive emotions such as \emoanticipation{anticipation} and \emojoy{joy} are less pronounced. Again, both platforms reacted to the unfolding events; however, we observe a less pronounced reaction on Parler.

\textbf{Derived emotions:} Compared to Twitter, we observe significantly higher levels of \emodisapproval{disapproval}, \emooutrage{outrage} and \emounbelief{unbelief} together with a lower level of \emoguilt{guilt} on Parler. For each of these emotions, KS-tests confirm that the differences in the distributions are statistically significant ($p<0.001$). Similar to our analysis of sentiment, we again observe a delayed response to the incidents on Parler. \normalcolor

\begin{figure}[ht]
    \centering
    \includegraphics[width=0.9\linewidth]{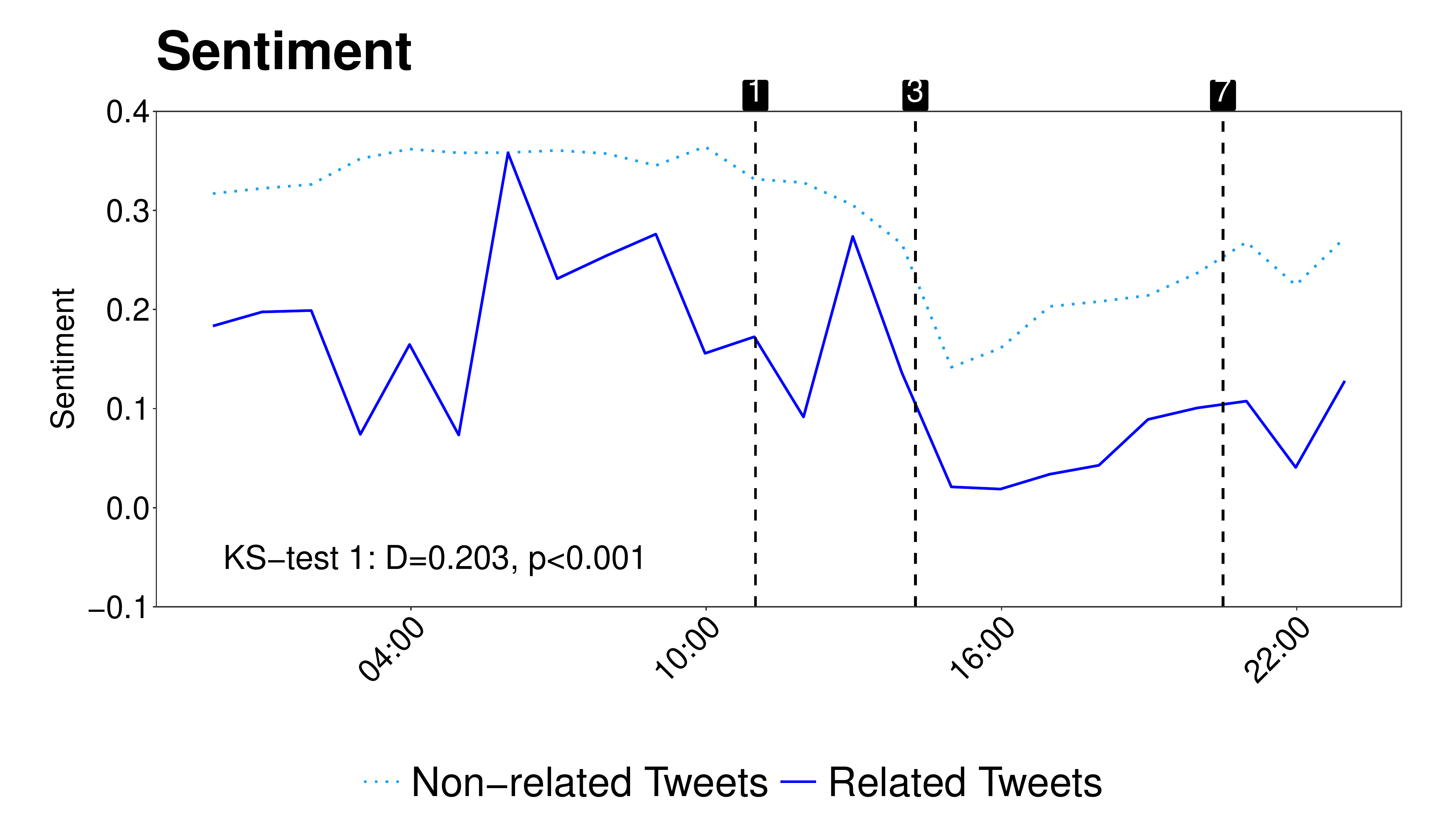}\\
    \caption{
    Sentiment (hourly) for storming-related and non-storming-related content on Twitter on January~6, 2021.}
    \label{fig:tweet_deepdive_1}
\end{figure}

To investigate the underlying sources for the dynamics, Fig.~\ref{fig:tweet_deepdive_1} divides the Twitter data into two subsets: (i)~tweets that contain keywords directly related to the storming of the \US capitol (\ie, ``riot'', ``rally'', ``Capitol'', ``speech'', ``POTUS'', ``America'', ``God bless'', `civil war''); and (ii) tweets without these keywords. We observe a significantly more negative sentiment in tweets related to the storming (KS-test: $D=0.203$, $p<0.001$). More importantly, during the time when rioters first entered the Capitol (Event \numbersymbol{3}), the drop in sentiment in these tweets is more pronounced (compared to unrelated tweets). These findings add to the validity of our results and suggest that the storming was responsible for the sentiment drop on Twitter.\normalcolor

\subsection{Robustness Checks}

We perform robustness checks to validate our results. (1)~Our dictionary approach based on the NRC was chosen due to its granularity, as well as the scarcity of other approaches (e.g., machine learning) that are similarly comprehensive \cite{Mohammad.2021}. Nevertheless, we find consistent results when repeating our analysis with different emotion dictionaries (\ie, NRC emotion lexicon, Linguistic Inquiry and Word Count (LIWC) emotion lexicon). (2)~Negations invert the meaning of Parleys. Our validation study with human annotations suggests that our approach correlates with human ratings. As a check, we used word lists to invert the valence of emotional words after the occurrence of tokens from that word list \citep{EPJ.2021}, which led to consistent findings. (3)~We repeated our experiments on emotional dynamics across user groups without overlapping users. Here, our findings remain consistent when excluding overlapping users from our analysis. (4)~We also repeated our analysis for larger time periods. We observe, for instance, a particularly positive sentiment and particularly positive emotions (\eg, \emojoy{joy}) at New Year's Eve (cf. Fig.~\ref{fig:time_series_sentiment}), while the most negative sentiment, as well as the largest share of negative emotions (\eg, \emofear{fear}, \emodisapproval{disapproval}), are still found during the storming of the U.S. Capitol. 

\section{Discussion}

\subsection{Relevance}

Understanding online diffusion on Parler is relevant due to the platforms' unique role that sets it aside from mainstream social media. Parler is highly popular among conservatives, right-wing extremists, and conspiracy theorists \cite{Romero.12.01.2021,Baer.2022}. Our analysis focused on the emotional reactions of users that self-identify with certain groups (Trump supporters, QAnon supporters, and supporters of alleged fraud). These small but particularly vocal groups (\eg in aggregate more than 10,000 posts on January~6, 2021, see Table~\ref{tab:summary-usergroup}) were characterized by significant differences in their (emotional) reactions to the violent incidents on January 6, 2021. In this regard, our results are also important as to whether social media platforms need to take a stronger stand against such user groups.

Dissecting the role of emotions in online diffusion is relevant for several reasons. Emotions have been found to influence user behavior \cite{Luminet.2000,Peters.2009}. Negative emotions are largely responsible for retweet behavior and influence which posts become viral \cite{Stieglitz.2013,Vosoughi.2018}. Hence, negative emotions (\eg, \emooutrage{outrage}) can contribute to a proliferation of conspiracy theories. Emotions are further highly contagious \cite{Kramer.2014}. This has alarming implications: If negative emotions proliferate, it is likely that other users, even outside of right-wing communities, may have also adopted them in their emotional state.

Emotions also play a central role in protests and allow for critical insights into their dynamics. Here, emotions can determine the interaction of protesters and opponents \citep{VanTroost.2013}. Research further suggests that people can evaluate the same event differently dependent on their group membership -- and consequently have different emotional reactions \citep{VanTroost.2013}. It is thus important to capture emotion dynamics across different user groups in order to better understand the role of emotions as drivers of protests. It may also allow us to speculate that the event could have been foreseen, alone by monitoring online emotions.

\subsection{Interpretation}

Our work finds distinctive emotional dynamics during the storming of the Capitol on Parler:
\begin{enumerate}
\item The Parler network embedded a more negative sentiment during and after the storming. This is accompanied by higher levels of \emosurprise{surprise}, \emofear{fear}, \emodisapproval{disapproval}, \emounbelief{unbelief}, and \emooutrage{outrage}. These findings are in line with research on emotions in protests, where, for example, \emooutrage{outrage} is considered a reactive emotion, triggered by activists toward concrete policies and decision-makers \citep{Jasper.2018}. Moreover, \emooutrage{outrage} is regarded as especially influential in online environments \citep{Crockett.2017}.
\item Trump supporters responded differently: they embedded a more negative sentiment during and after the storming than compared before the storming. They were also characterized by higher levels of \emodisapproval{disapproval} and \emooutrage{outrage} during the storming. After the storming, Parleys from Trump supporters embedded higher levels of \emounbelief{unbelief}.
\item QAnon supporters did not express a more negative sentiment during the storming. Instead, the sentiment remained relatively constant before, during, and after the storming. In contrast to other subgroups and the overall network, the emotional states of \emodisapproval{disapproval}, \emooutrage{outrage}, and \emoguilt{guilt} did not significantly vary on January~6. This is concerning as \emoguilt{guilt} is an important emotion after protests, which, when present, can induce an individual, for example, to not transgress social norms and values in the future \citep{VanTroost.2013}. These findings align with the general notion that QAnon supporters have questionable attitudes toward violence and democracy \cite{Papasavva.2021}.
\item Parler and Twitter both responded with a negative sentiment to the storming of the U.S. Capitol. However, the (negative) response on Parler was less pronounced compared to Twitter and lagged behind the (negative) response on Twitter. Moreover, compared to Twitter, emotions like \emodisapproval{disapproval} and \emooutrage{outrage} were significantly more pronounced on Parler. Overall, this suggests that Twitter embedded a negative sentiment towards the storming itself, while the sentiment on Parler was significantly lower also in the time before the storming, which may be attributed to different goals (\ie, contesting the election results). In addition, high levels of high-arousal negative emotions like \emooutrage{outrage} on Parler point towards higher affective polarization on Parler compared to Twitter.
\end{enumerate}
Consistent with research from political science \citep[\eg][]{VanTroost.2013}, we find that emotions are highly characteristic of riots. Moreover, we find that the emotional responses of social media users to the \emph{same} event can significantly vary depending on their group membership (\eg, Trump supporters vs. QAnon supporters). This observation is particularly important as it indicates that emotion dynamics are characteristic for specific user groups. Understanding these emotional dynamics may help to better anticipate the development of subgroups in online social media based on changes in emotion levels. Previous research on social media use around specific riots found a shift in content. For example, an increase of alt-right topics on Gab has been observed in the time before the Pittsburgh Synagogue shooting \citep{McIlroy.2019}. However, alt-tech social media platforms such as Parler are already characterized by relatively homogeneous content (compared to mainstream platforms), implying that not only the content itself but also emotions play a decisive role before, during, and after riots.

\subsection{Implications}

Our findings have several important implications. For social media platforms, questions arise about how similar events can be prevented. Online emotions appear to be deeply linked with the storming. Platforms should seek solutions based on which emotions can be monitored and actively managed to counter the proliferation of harmful content. For policy-makers, a better understanding of alt-tech and its societal implications is needed. This could result in tools for social media surveillance that eventually act as early warning systems for polarization tendencies. For research, our findings encourage granular assessments of online emotions \citep{Upworthy.2022}, where insights are generated not only for basic emotions but also for complex emotions such as \emodisapproval{disapproval}, \emounbelief{unbelief}, \emooutrage{outrage}, and \emoguilt{guilt}.

From a broader societal perspective, our findings shed new light on segregation and radicalization in the context of right-wing groups in the \US While the user base of alt-tech platforms such as Parler has commonly been viewed as being largely homogeneous, our analysis suggests that different sub-communities with varying motivations and goals have emerged. In particular, we find statistically significant differences in the emotional reactions to the violent incidents on January 6, 2021, across extremists users (QAnon supporters) and other, less radical user groups (\eg, Trump supporters). From a policy view, this finding is encouraging as it suggests that alt-tech platforms host a user base that -- at least in parts -- condemns violence and may still be accessible for (political) initiatives targeted at restoring social cohesion.

\textbf{Limitations:} As with other studies, our work is not free of limitations. (1)~Our results are limited by the accuracy of affective computing. To address this, we followed best-practice \citep{Song.2020} and performed a validation study with human ratings, finding a significant correlation. Future research may expand ours through more sophisticated, machine learning-based approaches for handling negations, irony, and satire. (2)~Our main aim is to quantify the use of emotional language of senders. As such, we do not attempt to infer the internal state, \ie, the feelings, of readers \citep{Kross.2019}. Nevertheless, our qualitative check using external labelers finds such correlation. (3)~Our analysis focused on users that self-identify with certain groups (\ie, users who are engaged), which reduces risks of type-II errors and is ought to yield more reliable inferences. (4)~We focused on specific emotions such as outrage and guilt due to their importance for protests \citep{Flam.2007,Jasper.2018,VanTroost.2013}. For this reason, we based our analysis on the NRC emotion lexicon due to its comprehensiveness \citep{Mohammad.2021}. Nevertheless, we acknowledge that there are other emotion models.

\section{Conclusion}

The social media platform Parler was used extensively during the storming of the U.S. Capitol, for which we examine the online emotions in this study. Specifically, we analyzed 717,300 posts that originated from 144,003 users and, using affective computing, provided a comprehensive assessment across different emotions. We found that the network content during the storming was characterized by an overall negative sentiment, with substantial heterogeneity across time and user groups. 

\subsection{Ethics statement}
We respect the privacy and agency of all people potentially impacted by this work and take specific steps to protect their privacy (see main text). The analysis was conducted in accordance with the Institutional Review Board at ETH Zurich.

\bibliography{references}

\begin{thebibliography}{62}
\providecommand{\natexlab}[1]{#1}
\providecommand{\url}[1]{\texttt{#1}}
\providecommand{\urlprefix}{URL }
\expandafter\ifx\csname urlstyle\endcsname\relax
  \providecommand{\doi}[1]{doi:\discretionary{}{}{}#1}\else
  \providecommand{\doi}{doi:\discretionary{}{}{}\begingroup
  \urlstyle{rm}\Url}\fi

\bibitem[{{ABC News}(2021)}]{Romero.12.01.2021}
{ABC News}. 2021.
\newblock Experts say echo chambers from apps like {Parler} and {Gab}
  contributed to attack on {Capitol}.
\newblock
  \urlprefix\url{https://abcnews.go.com/US/experts-echo-chambers-apps-parler-Gab-contributed-attack/story?id=75141014}.

\bibitem[{Aliapoulios et~al.(2021)Aliapoulios, Bevensee, Blackburn, Bradlyn,
  de~Cristofaro, Stringhini, and Zannettou}]{Aliapoulios.2021}
Aliapoulios, M.; Bevensee, E.; Blackburn, J.; Bradlyn, B.; de~Cristofaro, E.;
  Stringhini, G.; and Zannettou, S. 2021.
\newblock A large open dataset from the {Parler} social network.
\newblock In \emph{ICWSM}.

\bibitem[{Bakshy, Messing, and Adamic(2015)}]{Bakshy.2015}
Bakshy, E.; Messing, S.; and Adamic, L.~A. 2015.
\newblock Exposure to ideologically diverse news and opinion on {F}acebook.
\newblock \emph{Science} 348(6239): 1130--1132.

\bibitem[{{BBC News}(2021)}]{BBC.2021}
{BBC News}. 2021.
\newblock {US} {Capitol} riots: World leaders react to 'horrifying' scenes in
  Washington.
\newblock \urlprefix\url{https://www.bbc.com/news/world-us-canada-55568613}.

\bibitem[{Brady et~al.(2017)Brady, Wills, Jost, Tucker, and {van
  Bavel}}]{Brady.2017}
Brady, W.~J.; Wills, J.~A.; Jost, J.~T.; Tucker, J.~A.; and {van Bavel}, J.~J.
  2017.
\newblock Emotion shapes the diffusion of moralized content in social networks.
\newblock \emph{PNAS} 114(28): 7313--7318.

\bibitem[{Bär, Pröllochs, and Feuerriegel(2022)}]{Baer.2022}
Bär, D.; Pröllochs, N.; and Feuerriegel, S. 2022.
\newblock Finding Qs: Profiling QAnon supporters on Parler.
\newblock \emph{arXiv} .

\bibitem[{Conover et~al.(2011)Conover, Ratkiewicz, Francisco, Goncalves,
  Menczer, and Flammini}]{MichaelConover.2011}
Conover, M.; Ratkiewicz, J.; Francisco, M.; Goncalves, B.; Menczer, F.; and
  Flammini, A. 2011.
\newblock Political polarization on {Twitter}.
\newblock In \emph{ICWSM}.

\bibitem[{Crockett(2017)}]{Crockett.2017}
Crockett, M.~J. 2017.
\newblock Moral outrage in the digital age.
\newblock \emph{Nature Human Behaviour} 1(11): 769--771.

\bibitem[{{Daily Mail}(2021)}]{Jewers.2021}
{Daily Mail}. 2021.
\newblock Where {Capitol} rioters were when they posted gloating {Parler} siege
  videos: Hack of banned right-wing app geo-locates.
\newblock
  \urlprefix\url{https://www.dailymail.co.uk/news/article-9142985/GPS-data-Parler-hack-shows-users-app-stormed-Capitol
  -Building.html}.

\bibitem[{dos Reis~Costa et~al.(2015)dos Reis~Costa, Rotabi, Murnane, and
  Choudhury}]{Costa.2015}
dos Reis~Costa, J.~M.; Rotabi, R.; Murnane, E.~L.; and Choudhury, T. 2015.
\newblock It is not only about grievances: Emotional dynamics in social media
  during the Brazilian protests.
\newblock In \emph{ICWSM}.

\bibitem[{Druckman and Levendusky(2019)}]{Druckman.2019}
Druckman, J.~N.; and Levendusky, M.~S. 2019.
\newblock What Do We Measure When We Measure Affective Polarization?
\newblock \emph{Public Opinion Quarterly} 83(1): 114--122.

\bibitem[{Ekman(1992)}]{Ekman.1992}
Ekman, P. 1992.
\newblock An argument for basic emotions.
\newblock \emph{Cognition {\&} Emotion} 6(3-4): 169--200.

\bibitem[{Finkel et~al.(2020)Finkel, Bail, Cikara, Ditto, Iyengar, Klar, Mason,
  McGrath, Nyhan, Rand et~al.}]{Finkel.2020}
Finkel, E.~J.; Bail, C.~A.; Cikara, M.; Ditto, P.~H.; Iyengar, S.; Klar, S.;
  Mason, L.; McGrath, M.~C.; Nyhan, B.; Rand, D.~G.; et~al. 2020.
\newblock Political sectarianism in America.
\newblock \emph{Science} 370(6516): 533--536.

\bibitem[{Flam(2007)}]{Flam.2007}
Flam, H. 2007.
\newblock Emotions’ map: A research agenda.
\newblock In \emph{Emotions and social movements}, 29--50. Routledge.

\bibitem[{{Fortune}(2020)}]{Pressman.29.06.2020}
{Fortune}. 2020.
\newblock {Parler} is the new {Twitter} for conservatives. Here's what you need
  to know.
\newblock
  \urlprefix\url{https://fortune.com/2020/06/29/what-is-parler-app-social-media-conservatives-who-owns-free-echo-facebook-twitter-verified-faq/}.

\bibitem[{Freelon, Marwick, and Kreiss(2020)}]{Freelon.2020}
Freelon, D.; Marwick, A.; and Kreiss, D. 2020.
\newblock False equivalencies: Online activism from left to right.
\newblock \emph{Science} 369(6508): 1197--1201.

\bibitem[{Goodwin, Jasper, and Polletta(2009)}]{Goodwin.2009}
Goodwin, J.; Jasper, J.~M.; and Polletta, F. 2009.
\newblock \emph{Passionate politics: Emotions and social movements}.
\newblock University of Chicago Press.

\bibitem[{Grinberg et~al.(2019)Grinberg, Joseph, Friedland, Swire-Thompson, and
  Lazer}]{Grinberg.2019}
Grinberg, N.; Joseph, K.; Friedland, L.; Swire-Thompson, B.; and Lazer, D.
  2019.
\newblock Fake news on {T}witter during the 2016 {U.S.} presidential election.
\newblock \emph{Science} 363(6425): 374--378.

\bibitem[{Hitkul et~al.(2021)Hitkul, Prabhu, Guhathakurta, jain, Subramanian,
  Reddy, Sehgal, Karandikar, Gulati, Arora, Shah, and Kumaraguru}]{Hitkul.2021}
Hitkul; Prabhu, A.; Guhathakurta, D.; jain, J.; Subramanian, M.; Reddy, M.;
  Sehgal, S.; Karandikar, T.; Gulati, A.; Arora, U.; Shah, R.~R.; and
  Kumaraguru, P. 2021.
\newblock {Capitol} (Pat)riots: A comparative study of {Twitter} and {Parler}.
\newblock \emph{arXiv} 2101.06914.

\bibitem[{Iyengar et~al.(2019)Iyengar, Lelkes, Levendusky, Malhotra, and
  Westwood}]{Iyengar.2019}
Iyengar, S.; Lelkes, Y.; Levendusky, M.; Malhotra, N.; and Westwood, S.~J.
  2019.
\newblock The origins and consequences of affective polarization in the {United
  States}.
\newblock \emph{Annual Review of Political Science} 22(1): 129--146.

\bibitem[{Jacoby(2000)}]{Jacoby.2000}
Jacoby, W.~G. 2000.
\newblock Loess.
\newblock \emph{Electoral Studies} 19(4): 577--613.

\bibitem[{Jacomy et~al.(2014)Jacomy, Venturini, Heymann, and
  Bastian}]{Jacomy.2014}
Jacomy, M.; Venturini, T.; Heymann, S.; and Bastian, M. 2014.
\newblock ForceAtlas2, a continuous graph layout algorithm for handy network
  visualization designed for the Gephi software.
\newblock \emph{PLOS ONE} 9(6): e98679.

\bibitem[{Jasper(1998)}]{Jasper.1998}
Jasper, J.~M. 1998.
\newblock The emotions of protest: Affective and reactive emotions in and
  around social movements.
\newblock In \emph{Sociological Forum}, volume~13, 397--424.

\bibitem[{Jasper(2018)}]{Jasper.2018}
Jasper, J.~M. 2018.
\newblock \emph{The emotions of protest}.
\newblock University of Chicago Press.

\bibitem[{Kim and Yoo(2012)}]{Kim.2012}
Kim, J.; and Yoo, J. 2012.
\newblock Role of sentiment in message propagation: Reply vs. retweet behavior
  in political communication.
\newblock In \emph{Social Informatics}, 131--136.

\bibitem[{Kramer, Guillory, and Hancock(2014)}]{Kramer.2014}
Kramer, A. D.~I.; Guillory, J.~E.; and Hancock, J.~T. 2014.
\newblock Experimental evidence of massive-scale emotional contagion through
  social networks.
\newblock \emph{PNAS} 111(24): 8788--8790.

\bibitem[{Kratzwald et~al.(2018)Kratzwald, Ili{\'c}, Kraus, Feuerriegel, and
  Prendinger}]{Kratzwald.2018}
Kratzwald, B.; Ili{\'c}, S.; Kraus, M.; Feuerriegel, S.; and Prendinger, H.
  2018.
\newblock Deep learning for affective computing: Text-based emotion recognition
  in decision support.
\newblock \emph{Decision Support Systems} 115: 24--35.

\bibitem[{Kross et~al.(2019)Kross, Verduyn, Boyer, Drake, Gainsburg, Vickers,
  Ybarra, and Jonides}]{Kross.2019}
Kross, E.; Verduyn, P.; Boyer, M.; Drake, B.; Gainsburg, I.; Vickers, B.;
  Ybarra, O.; and Jonides, J. 2019.
\newblock Does counting emotion words on online social networks provide a
  window into people’s subjective experience of emotion? A case study on
  Facebook.
\newblock \emph{Emotion} 19(1): 97.

\bibitem[{Lazer et~al.(2018)Lazer, Baum, Benkler, Berinsky, Greenhill, Menczer,
  Metzger, Nyhan, Pennycook, Rothschild, Schudson, Sloman, Sunstein, Thorson,
  Watts, and Zittrain}]{Lazer.2018}
Lazer, D. M.~J.; Baum, M.~A.; Benkler, Y.; Berinsky, A.~J.; Greenhill, K.~M.;
  Menczer, F.; Metzger, M.~J.; Nyhan, B.; Pennycook, G.; Rothschild, D.;
  Schudson, M.; Sloman, S.~A.; Sunstein, C.~R.; Thorson, E.~A.; Watts, D.~J.;
  and Zittrain, J.~L. 2018.
\newblock The science of fake news.
\newblock \emph{Science} 359(6380): 1094--1096.

\bibitem[{Luminet et~al.(2000)Luminet, Bouts, Delie, Manstead, and
  Rim{\'e}}]{Luminet.2000}
Luminet, O.; Bouts, P.; Delie, F.; Manstead, A. S.~R.; and Rim{\'e}, B. 2000.
\newblock Social sharing of emotion following exposure to a negatively valenced
  situation.
\newblock \emph{Cognition {\&} Emotion} 14(5): 661--688.

\bibitem[{Lutz et~al.(2020)Lutz, Adam, Feuerriegel, Pr{\"o}llochs, and
  Neumann}]{Lutz.2020}
Lutz, B.; Adam, M. T.~P.; Feuerriegel, S.; Pr{\"o}llochs, N.; and Neumann, D.
  2020.
\newblock Affective information processing of fake news: Evidence from
  {NeuroIS}.
\newblock In \emph{Lecture Notes in Information Systems and Organisation},
  volume~32, 121--128.

\bibitem[{McIlroy-Young and Anderson(2019)}]{McIlroy.2019}
McIlroy-Young, R.; and Anderson, A. 2019.
\newblock {From “welcome new gabbers” to the {Pittsburgh synagogue}
  shooting: The evolution of gab}.
\newblock In \emph{ICWSM}.

\bibitem[{Mohammad(2021)}]{Mohammad.2021}
Mohammad, S.~M. 2021.
\newblock Sentiment analysis: Automatically detecting valence, emotions, and
  other affectual states from text.
\newblock In \emph{Emotion Measurement}, 323--379.

\bibitem[{Munn(2021)}]{Munn.2021}
Munn, L. 2021.
\newblock More than a mob: Parler as preparatory media for the U.S. Capitol
  storming.
\newblock \emph{First Monday} 26(3).

\bibitem[{Naumzik and Feuerriegel(2022)}]{Naumzik.2022}
Naumzik, C.; and Feuerriegel, S. 2022.
\newblock Detecting false rumors from retweet dynamics on social media.
\newblock In \emph{WWW}.

\bibitem[{Ng, Cruickshank, and Carley(2021)}]{Ng.2021}
Ng, L. H.~X.; Cruickshank, I.; and Carley, K.~M. 2021.
\newblock Coordinating narratives and the {Capitol} riots on {Parler}.
\newblock \emph{arXiv} 2109.00945.

\bibitem[{Otala et~al.(2021)Otala, Kurtic, Grasso, Liu, Matthews, and
  Madraki}]{Otala.2021}
Otala, J.~M.; Kurtic, G.; Grasso, I.; Liu, Y.; Matthews, J.; and Madraki, G.
  2021.
\newblock Political Polarization and Platform Migration.
\newblock In \emph{WWW}.

\bibitem[{Papasavva et~al.(2021)Papasavva, Blackburn, Stringhini, Zannettou,
  and Cristofaro}]{Papasavva.2021}
Papasavva, A.; Blackburn, J.; Stringhini, G.; Zannettou, S.; and Cristofaro,
  E.~D. 2021.
\newblock ``{Is} it a {Q}oincidence?'': {An} exploratory study of {QAnon} on
  {Voat}.
\newblock In \emph{WWW}.

\bibitem[{Peters, Kashima, and Clark(2009)}]{Peters.2009}
Peters, K.; Kashima, Y.; and Clark, A. 2009.
\newblock Talking about others: Emotionality and the dissemination of social
  information.
\newblock \emph{European Journal of Social Psychology} 39(2): 207--222.

\bibitem[{{Pew Research Center}(2020{\natexlab{a}})}]{Mitchell.2020}
{Pew Research Center}. 2020{\natexlab{a}}.
\newblock Americans who mainly get their news on social media are less engaged,
  less knowledgeable.
\newblock
  \urlprefix\url{https://www.journalism.org/2020/07/30/americans-who-mainly-get-their-news-on-social-media-are-less-engaged-less-knowledgeable/}.

\bibitem[{{Pew Research Center}(2020{\natexlab{b}})}]{Schaeffer.2020}
{Pew Research Center}. 2020{\natexlab{b}}.
\newblock Fast facts about Americans' views of social media companies as
  {Trump}-{Twitter} dispute grows.
\newblock
  \urlprefix\url{https://www.pewresearch.org/fact-tank/2020/05/29/fast-facts-about-americans-views-of-social-media-companies-as-Trump-twitter-dispute-grows/}.

\bibitem[{Plutchik(2001)}]{Plutchik.2001}
Plutchik, R. 2001.
\newblock The nature of emotions: Human emotions have deep evolutionary roots,
  a fact that may explain their complexity and provide tools for clinical
  practice.
\newblock \emph{American Scientist} 89(4): 344--350.

\bibitem[{Pr{\"o}llochs, B{\"a}r, and
  Feuerriegel(2021{\natexlab{a}})}]{ScientificReports.2021}
Pr{\"o}llochs, N.; B{\"a}r, D.; and Feuerriegel, S. 2021{\natexlab{a}}.
\newblock Emotions explain differences in the diffusion of true vs. false
  social media rumors.
\newblock \emph{Scientific Reports} 11(22721).

\bibitem[{Pr{\"o}llochs, B{\"a}r, and
  Feuerriegel(2021{\natexlab{b}})}]{EPJ.2021}
Pr{\"o}llochs, N.; B{\"a}r, D.; and Feuerriegel, S. 2021{\natexlab{b}}.
\newblock Emotions in online rumor diffusion.
\newblock \emph{EPJ Data Science} 10(1).

\bibitem[{ProPublica(2021)}]{ProPublica.2021}
ProPublica. 2021.
\newblock What {Parler} Saw During the Attack on the {Capitol}.
\newblock
  \urlprefix\url{https://projects.propublica.org/parler-capitol-videos/}.

\bibitem[{Recuero, Soares, and Gruzd(2020)}]{Recuero.2020}
Recuero, R.; Soares, F.~B.; and Gruzd, A. 2020.
\newblock Hyperpartisanship, Disinformation and Political Conversations on
  {Twitter}: The Brazilian Presidential Election of 2018.
\newblock In \emph{ICWSM}.

\bibitem[{Rivers and Lewis(2014)}]{Rivers.2014}
Rivers, C.~M.; and Lewis, B.~L. 2014.
\newblock Ethical research standards in a world of big data.
\newblock \emph{F1000Research} 3: 38.

\bibitem[{Robertson et~al.(2022)Robertson, Pröllochs, Schwarzenegger,
  Parnamets, Van~Bavel, and Feuerriegel}]{Upworthy.2022}
Robertson, C.; Pröllochs, N.; Schwarzenegger, K.; Parnamets, P.; Van~Bavel,
  J.~J.; and Feuerriegel, S. 2022.
\newblock Negativity drives online news consumption.
\newblock
  \urlprefix\url{https://figshare.com/articles/journal_contribution/Negativity_drives_online_news_consumption_Registered_Report_Stage_1_Protocol_/19657452}.

\bibitem[{Sauter et~al.(2010)Sauter, Eisner, Ekman, and Scott}]{Sauter.2010}
Sauter, D.~A.; Eisner, F.; Ekman, P.; and Scott, S.~K. 2010.
\newblock Cross-cultural recognition of basic emotions through nonverbal
  emotional vocalizations.
\newblock \emph{PNAS} 107(6): 2408--2412.

\bibitem[{Solovev and Pr{\"o}llochs(2022)}]{Solovev.2022}
Solovev, K.; and Pr{\"o}llochs, N. 2022.
\newblock Moral emotions shape the virality of COVID-19 misinformation on
  social media.
\newblock In \emph{WWW}.

\bibitem[{Song et~al.(2020)Song, Tolochko, Eberl, Eisele, Greussing,
  Heidenreich, Lind, Galyga, and Boomgaarden}]{Song.2020}
Song, H.; Tolochko, P.; Eberl, J.-M.; Eisele, O.; Greussing, E.; Heidenreich,
  T.; Lind, F.; Galyga, S.; and Boomgaarden, H.~G. 2020.
\newblock In validations we trust? The impact of imperfect human annotations as
  a gold standard on the quality of validation of automated content analysis.
\newblock \emph{Political Communication} 37(4): 550--572.

\bibitem[{Starbird(2017)}]{Starbird.2017}
Starbird, K. 2017.
\newblock Examining the alternative media ecosystem through the production of
  alternative narratives of mass shooting events on {T}witter.
\newblock In \emph{ICWSM}.

\bibitem[{Stieglitz and Dang-Xuan(2013)}]{Stieglitz.2013}
Stieglitz, S.; and Dang-Xuan, L. 2013.
\newblock Emotions and information diffusion in social media: Sentiment of
  microblogs and sharing behavior.
\newblock \emph{JMIS} 29(4): 217--248.

\bibitem[{{The New York Times}(2021{\natexlab{a}})}]{Frenkel.06.01.2021}
{The New York Times}. 2021{\natexlab{a}}.
\newblock How The Storming of {Capitol} Hill Was Organized on Social Media.
\newblock
  \urlprefix\url{https://www.nytimes.com/2021/01/06/us/politics/protesters-storm-capitol-hill-building.html}.

\bibitem[{{The New York Times}(2021{\natexlab{b}})}]{NewYorkTimes.04.03.2021}
{The New York Times}. 2021{\natexlab{b}}.
\newblock What Is {QAnon}, the Viral Pro-{Trump} Conspiracy Theory?
\newblock \urlprefix\url{https://www.nytimes.com/article/what-is-qanon.html}.

\bibitem[{{The Wall Street Journal}(2021)}]{Wells.08.01.2021}
{The Wall Street Journal}. 2021.
\newblock `{Trump} or War': How the {Capitol} Mob Mobilized on Social Media.
\newblock
  \urlprefix\url{https://www.wsj.com/articles/Trump-or-war-how-the-capitol-mob-mobilized-on-social-media-11610069778}.

\bibitem[{Van~Troost, Van~Stekelenburg, and Klandermans(2013)}]{VanTroost.2013}
Van~Troost, D.; Van~Stekelenburg, J.; and Klandermans, B. 2013.
\newblock Emotions of protest.
\newblock In \emph{Emotions in Politics}, 186--203. Springer.

\bibitem[{Venkatesan et~al.(2021)Venkatesan, Valecha, Yaraghi, Oh, and
  Rao}]{Venkatesan.2021}
Venkatesan, S.; Valecha, R.; Yaraghi, N.; Oh, O.; and Rao, H.~R. 2021.
\newblock Influence in social media: An investigation of tweets spanning the
  2011 Egyptian revolution.
\newblock \emph{MIS Quarterly} 45(4): 1679--1714.

\bibitem[{Vosoughi(2015)}]{Vosoughi.2015}
Vosoughi, S. 2015.
\newblock \emph{Automatic detection and verification of rumors on {T}witter}.
\newblock Ph.D. thesis, {MIT}.

\bibitem[{Vosoughi, Roy, and Aral(2018)}]{Vosoughi.2018}
Vosoughi, S.; Roy, D.; and Aral, S. 2018.
\newblock The spread of true and false news online.
\newblock \emph{Science} 359(6380): 1146--1151.

\bibitem[{Zhang(2013)}]{Zhang.2013}
Zhang, P. 2013.
\newblock The affective response model: A theoretical framework of affective
  concepts and their relationships in the {ICT} context.
\newblock \emph{MIS Quarterly} 37(1): 247--274.

\bibitem[{Zollo et~al.(2015)Zollo, Novak, {Del Vicario}, Bessi, Mozeti{\v{c}},
  Scala, Caldarelli, and Quattrociocchi}]{Zollo.2015}
Zollo, F.; Novak, P.~K.; {Del Vicario}, M.; Bessi, A.; Mozeti{\v{c}}, I.;
  Scala, A.; Caldarelli, G.; and Quattrociocchi, W. 2015.
\newblock Emotional dynamics in the age of misinformation.
\newblock \emph{PLOS ONE} 10(9): e0138740.

\end{thebibliography}

\end{document}